\documentclass[twocolumn]{aastex631}
\usepackage[caption=false]{subfig}
\usepackage{graphicx}
\usepackage{units}
\usepackage{float}

\usepackage{color}

\definecolor{mygreen}{RGB}{84,141,40}

\begin{document}

\title{The Rapid Optical Variability of the Nearby Radio-Loud AGN Pictor~A: Introducing the \emph{Quaver} Pipeline for AGN Science with TESS}

\author[0000-0001-5785-7038]{Krista Lynne Smith}
\affiliation{Department of Physics and Astronomy, Texas A\&M University, College Station, TX 77845, USA}
\affiliation{Department of Physics, Southern Methodist University, 3215 Daniel Ave., Dallas, TX 75205, USA}

\author[0000-0001-8020-3884]{Lia F. Sartori}
\affiliation{Institute for Particle Physics and Astrophysics, ETH Z\"urich, Wolfgang-Pauli-Strasse 27, CH-8093 Z\"urich, Switzerland}

\begin{abstract}

The sampling strategy of the Transiting Exoplanet Survey Satellite (TESS) make TESS light curves extremely valuable to investigate high cadence optical variability of AGN. However, because the
TESS instrument was primarily designed for exoplanet
science, the use of the satellite for other applications requires
careful treatment of the data. In this paper we introduce \texttt{Quaver}, a new software tool designed specifically to extract TESS light curves of extended and faint sources presenting stochastic variability. We then use this new tool to extract light curves of the nearby radio-loud AGN Pictor~A, and perform a temporal and power spectral analysis of its high cadence optical variability.  The obtained light curves are well fit with a damped random walk (DRW) model, exhibiting both stochastic AGN variations and flaring behavior. The DRW characteristic timescales $\tau_{\rm DRW} \sim 3-6$ days during more quiet periods, and  $\tau_{\rm DRW} \sim 0.8$ days for periods with strong flares, even when the flares themselves are masked from the DRW fit. The observed timescales are consistent with the dynamical, orbital and thermal timescales expected for the low black hole mass of Pictor~A.

\end{abstract}


\section{Introduction} \label{sec:intro}
Optical variability of active galactic nuclei (AGN) provides one of the very few direct observational probes of physics within the accretion disk. The high-amplitude, coherent variations frequently seen in ground-based AGN optical monitoring projects are ascribed to a number of phenomena intrinsic to the accretion process, such as magnetic phenomena due to turbulence and reconnection (\citealt{Balbus1998}), the damped random walk of flux within the disk (e.g., \citealt{Kelly2009}), reprocessing of rapid X-ray variations in the corona, and overall variations in the mass accretion rate. 

 Approximately ten percent of AGN exhibit powerful jets observed primarily in the radio, often extending far beyond the extent of the host galaxy and considered an important driver of galaxy evolution via AGN feedback. Why some AGN are able to launch such jets while others do not remains an active area of research. The ability of an AGN to produce a jet may depend on the spin of the black hole \citep[e.g., ][]{Blandford1977,Wilson1995} or upon the accretion state of the system, analogous to XRBs (e.g., \citealt{Koerding2006}). 

With this in mind, it is important to study and compare the variability properties, and thus the accretion physics, of radio-loud and radio-quiet AGN. To this end, we have extracted and analyzed a high-cadence, high-precision optical light curve of the nearest broad-line radio galaxy (BLRG), Pictor~A (PKS~0518-548), from the Transiting Exoplanet Survey Satellite (TESS). This galaxy has a pronounced double-lobed radio structure known for decades (\citealt{Maltby1962}), and is the closest such source with a Type~1 optical spectrum at $z=0.035$. Studied extensively by \cite{Simkin1999}, Pictor~A has a compact, parsec-scale radio jet confirmed in VLBI imaging, and gas kinematics indicating a strong interaction of the radio jet with the interstellar medium. The galaxy is also a known gamma-ray emitter, with variable gamma-ray flux (\citealt{BrownAdams2012}), and X-ray knots co-spatial with radio knots in the jet, which also show strong variability (\citealt{Marshall2010}). The black hole mass in Pictor~A is likely quite low, similar to the mass of Sagittarius A$^*$, $M_{\rm BH}~=~5.9~\times~10^6 M_{\odot}$ \citep{Koss2022a}.

The TESS instrument \citep{Ricker2015}, designed primarily to search for transiting exoplanets using long-term monitoring with high photometric precision of stars across nearly the entire sky, provides optical monitoring at rapid cadence: every 30 minutes in the early cycles (2018 - 2019) and every 10 minutes or 2 minutes in later and current cycles (2020 - present). Because the spacecraft does not contend with diurnal or seasonal gaps, coverage is nearly continuous and much more complete than ground-based monitoring. 

These properties make TESS a potentially transformative instrument for timing of high-energy processes like accretion onto supermassive black holes, magnetic flares, and jetted phenomena. However, because the TESS instrument was primarily designed for exoplanet science, use of the satellite for other applications requires careful treatment of the data, as indicated in past studies of AGN using the very similar \emph{Kepler} satellite (e.g., \citealt{Kasliwal2015}, \citealt{Smith2018}, \citealt{Moreno2021}). 

In this paper, we introduce a new software tool designed specifically for sources with stochastic, or even non-stationary, variability and that can accommodate sources in extended host galaxies: the \texttt{Quaver} program. \texttt{Quaver} is designed to be fully interactive, transparent, and flexible; a ``Swiss army knife" for the extraction and correction of TESS light curves. Novel features include clickable interactive interfaces for the selection of the extraction aperture and masking of severely compromised cadences, and a sophisticated matrix regression method for the correction of background light and electronic systematics.

In the following document, we first discuss the TESS observations and describe the new software package (Section~\ref{sec:quaver}) and explain some caveats of the method (Section~\ref{sec:cautions}) before undertaking a temporal and power spectral analysis of the TESS light curve of Pictor~A as a demonstration of the software (Section~\ref{sec:vm}). We then discuss the implications of our results in Section~\ref{sec:disc}, before concluding with a summary in Section~\ref{sec:conclusion}. 

\section{TESS Data}\label{sec:tess_description}

The baseline for a TESS source depends upon its ecliptic latitude, ranging from 27 days near the ecliptic plane to approximately 1 year at the ecliptic poles. The main output product of TESS are the Full Frame Images (FFIs), from which the user extracts their own light curves. In the mission's Cycles~1 and 2 (July 2018-June 2020), the cadence of these FFIs was 30~minutes. In Cycles 3 and 4 (July 2020 - Aug 2022, the cadence was raised to 10~minutes), and in the current Cycle~5 and Cycle~6 (Sep 2022 - Sep 2024), the cadence is 200~seconds. The frequency of data downlinks has also increased from the earliest cycles, from 13.7~day intervals to 7~day intervals. TESS monitored the section of sky covering Pictor~A as part of its routine Cycle~1 survey operations between October~18 2018 and January 7 2019 with 30-minute cadence. This span of time consists of three separate sectors (Sectors~4, 5, and 6) of 27 days each, for a total baseline of 81 days. Each TESS sector consists of two orbits, in between which the satellite downlinks to Earth to transmit data; such a transmission also occurs between sectors. There is therefore a $\sim$1-day gap approximately every two weeks during the monitoring. Additionally, during Sector~4 there was an instrument anomaly that paused data collection for three days, and the first three days of Sector~6 were used for calibration with no science data collected; see the TESS Data Release Notes\footnote{\url{https://archive.stsci.edu/tess/tess_drn.html}} for detailed information. 

Pictor~A was further observed during Cycle~3 Sectors 31 and 33, from October 21 2020 to November 19 2020, and again from December 17 2020 to January 13 2021, with a 27-day gap in between. These observations had the new, faster Cycle~3 cadence of 10-minutes. A star-tracker anomaly caused Sector~31 to terminate science data collection two days early. 

The TESS bandpass is very wide, with no capability to collect light curves in different bands, similar to that of \emph{Kepler} \citep{Borucki2010}. In contrast to \emph{Kepler}'s ``white-light" bandpass, TESS instead spans a ``pink" wavelength range of 600-1000~nm to best encapsulate the spectral energy distribution of M-dwarfs.

\section{Data Reduction with \emph{Quaver}}\label{sec:quaver}
\subsection{Motivation}
\label{sec:motivation}
Data collected from satellites primarily concerned with detecting the periodic signal of exoplanet transits are not optimized for the study of stochastic variability in sources like AGN. In addition, the majority of detrending techniques used to reduce the data to search for periodic signals in planet hosts or astroseismic targets are inappropriate for treating AGN variability, frequently resulting in overfitting.

Problematically, many of the systematic trends in space-based optical light curves, in both the \emph{Kepler} and TESS data, can mimic true AGN variability and are difficult to remove using simple background-subtraction techniques. Furthermore, the fact that many AGN host galaxies are extended sources and not point sources like stars requires special attention to be paid to the extraction aperture, rather than relying on simple PSF modeling. 

\begin{figure}
\centering
\includegraphics[width=0.4\textwidth]{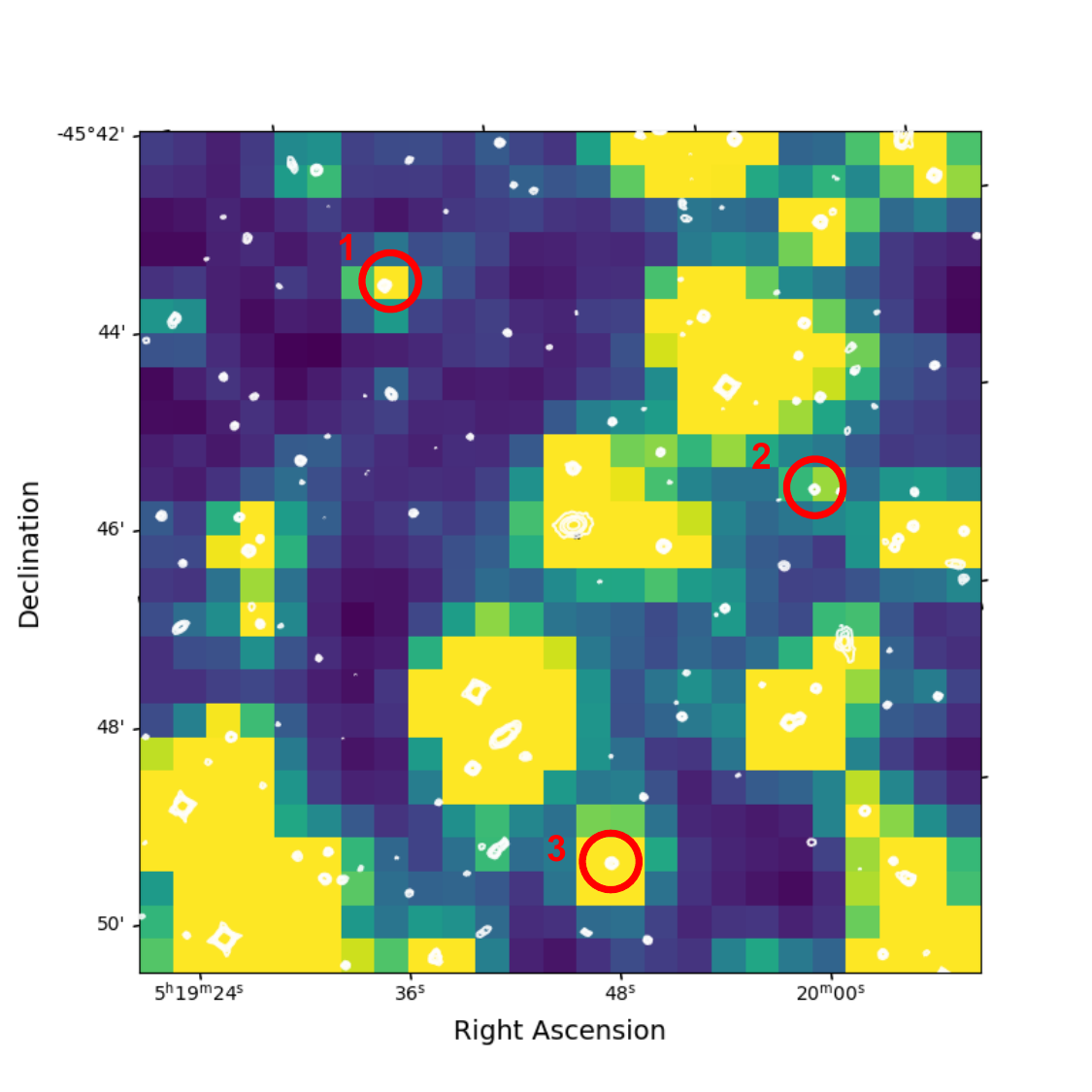}
\caption{TESS Target Pixel File image of the Pictor~A field of view, with contours from the DSS overlaid. The three reference stars extracted by the pipeline are labeled in red circles; their light curves are shown in Figure~\ref{fig:ref_stars_lcs}. The scale is 21\arcsec per pixel. Pictor~A is the galaxy at the center of the frame.}
\label{fig:ref_stars_map}
\end{figure}

In answer to these challenges, we have developed the software package \texttt{Quaver}, which we make publicly available\footnote{\url{https://github.com/kristalynnesmith/quaver}}. \texttt{Quaver} delivers an interactive, transparent, and customizeable TESS extraction procedure for quasar-like variability, but can be used for any type of source that may benefit from a tailored extraction approach. The software page linked below has a full user guide available, with walk-throughs of several source cases. While not yet practical for use on large numbers of sources that can be extracted using automated optimization procedures, \texttt{Quaver} is intended for sources in which a careful and deliberate approach is required, allowing the user to inspect the removed components and explore the effect of different correction techniques and apertures on the robustness of the extracted light curve.

\subsection{Aperture Selection}
\label{sec:aperture}
The extraction process begins with an interactive interface for choosing a custom extraction aperture by click-selecting desired pixels. This is an especially useful feature for AGN, which typically reside in extended host galaxies that may overlap in unpredictable ways with pre-set aperture shapes or complicate PSF-fitting techniques.

There are several effects that can cause the selected aperture to encompass more or less of the wings of the flux distribution as a function of time, introducing spurious variability in the light curves. These include: 
\begin{itemize}
    \item pointing jitter, which causes the distribution of the source PSF across the pixels and their different response functions to change;
    \item differential velocity aberration, an effect of the spacecraft's motion around the solar system barycenter that causes the source position to drift over a sector; and
    \item thermal breathing, in which the temperature of the optics changes during the orbit and causes the PSF to become larger or smaller.
\end{itemize} 

It is important to realize that although the TESS pixels are quite large, the PSF of even a point source typically extends into nearby pixels. Indeed, the PSF varies across the field of view of the detector, from $\sim0.9$~ pixels near the center to as much as 2.76 pixels near the edges \citep{Oelkers2018}. It is therefore recommended that the user attempt to encompass the \emph{entire host galaxy} in the aperture, if it is feasible. In fields that are not crowded, it can be useful to include a buffer pixel around one's source; however, for faint sources, this can introduce significant background noise that may dilute the light curve substantially. It is also important to realize that increasing the aperture size will dilute the variable signal from the nucleus relative to the baseline flux of the light curve, changing the overall percent variability.

In general, it is best to make the aperture sufficiently large to encompass the host galaxy flux and its likely spillover into nearby pixels, while not encompassing so much background as to raise the noise level of the light curves beyond what your desired variability threshold can handle. Experimentation with different apertures around extended sources is recommended: generally, a well-chosen aperture will result in an output light curve that retains its shape, and only grows more noisy, when additional pixels are added (assuming all other extraction parameters remain the same). An aperture that is too small will result in false variations when drift or thermal effects cause portions of the source flux distribution to enter or leave the aperture. An aperture that is too large will be excessively noisy, and of course, an aperture should not be so large as to include other sources.

The custom aperture also allows the user to avoid nearby contaminants or determine the variability of nearby stars or background regions. The user selects pixels for inclusion in the aperture from a clicakble map, generated by TESSCut \citep{Brasseur2019}.  The process is assisted by an overlay of contours from the Digitized Sky Survey (DSS) atop the very large 21\arcsec~ TESS pixels, so that the user can identify any potential contaminants and better assess the extent of the host galaxy. An image of the type of map shown during this selection is given in Figure~\ref{fig:ref_stars_map}.  
\subsection{Systematics Correction}
\label{sec:sys_corr}
Once the aperture is selected, the \texttt{Quaver} program utilizes both principal component analysis (PCA) and matrix regression to account for systematics. It makes frequent use of algorithms developed for \texttt{Lightkurve}, originally developed for the \emph{Kepler} mission and now adapted for TESS analysis \citep{Lightkurve2018}. 

\subsubsection{Classification of Pixels}
\label{sec:pixel_class}
Once the aperture is chosen, the pixel map is divided into three regions: the source aperture plus a 1-pixel border buffer zone (to prevent light from the source from leaking into pixels used to build correction vectors), ``faint" background pixels, and ``bright" pixels that are likely to contain other sources. This cut is made at 1.5$\sigma$ above the median flux of the whole pixel map by default, but can be changed by the user if desired. In a crowded field, the number of background pixels will be fewer than in a field that is mostly empty. If the field is very crowded, you may wish to increase the 1.5$\sigma$ threshold, otherwise several actual faint background sources may be included in the ``empty" pixels.

\subsubsection{Cadence Masking}
\label{sec:cadence_masking}
A principal component analysis is performed on the faint pixels. This next step allows the user to see what the dominant effects of this background look like. \texttt{Quaver} will plot the \textbf{eigenvalues} of the principal components of the faint pixels. The tunable parameter \texttt{sys\_threshold} sets the value that determines whether this plot appears; it is set at the arbitrary value of 0.2 by default, but can be altered. If any of the components exceeds this value at any point in the light curve, the user will be asked whether they wish to interactively mask out periods of monitoring. This is entirely optional. The user can then click-select regions to exclude from the light curve (and fitting) moving forward. This is frequently useful, as there are many TESS sectors where brief periods of spacecraft attitude corrections, erroneous pointing solutions, or necessary thruster firings cause rapid, erratic behavior that is very hard or impossible to remove. The beginnings of the two orbits in each sector are prone to thermal effects, as the spacecraft recovers equilibrium after data downlinks at orbit perigees; the beginnings and ends of orbits are especially prone to increased scattered light from the Earth. \texttt{Quaver} removes these gradual effects successfully in most cases; however, masking these edge cadences can be useful to prevent spurious behavior resembling partial flares or rising/falling flux near sector edges from remaining in the light curve. 

The user should think carefully about the effects of removing portions of the light curve with masking. For example:

\begin{itemize}
    \item Removing significant portions of the light curve edges reduces the overall baseline, shortening the frequency range to which the analysis is sensitive. For AGN, this can mean the inability to detect low-frequency turnovers.
    \item Removing various stretches of the light curve throughout renders the TESS sampling much more uneven, meaning that methods such as the Lomb-Scargle Periodogram are more appropriate than, say, interpolation over the orbit gap followed by simple Fourier power spectral analysis.
\end{itemize}

A conservative approach to masking is to decline the option to mask at first, and allow the program to try and correct the systematics. If the final output light curve remains visibly affected by a known issue from the Data Release Notes or serious thermal edge effects, then the program can be re-run with the masking option utilized.

\subsubsection{Matrix Regression Correction}
\label{sec:matrix_regression}
Once the aperture is chosen and any desired cadences are masked out, \texttt{Quaver} moves on to the systematics correction itself.
 
The \texttt{Quaver} pipeline has three options for systematics correction. In order of complexity and aggressiveness of fitting, they are: a simple PCA method, a simple hybrid method, and a fully hybrid method. The hybrid methods are so-named because they handle the additive background effects and the multiplicative instrumental effects separately, in a hybrid approach.


\begin{figure*}%
    \centering
    \subfloat
    {{\includegraphics[width=8cm]{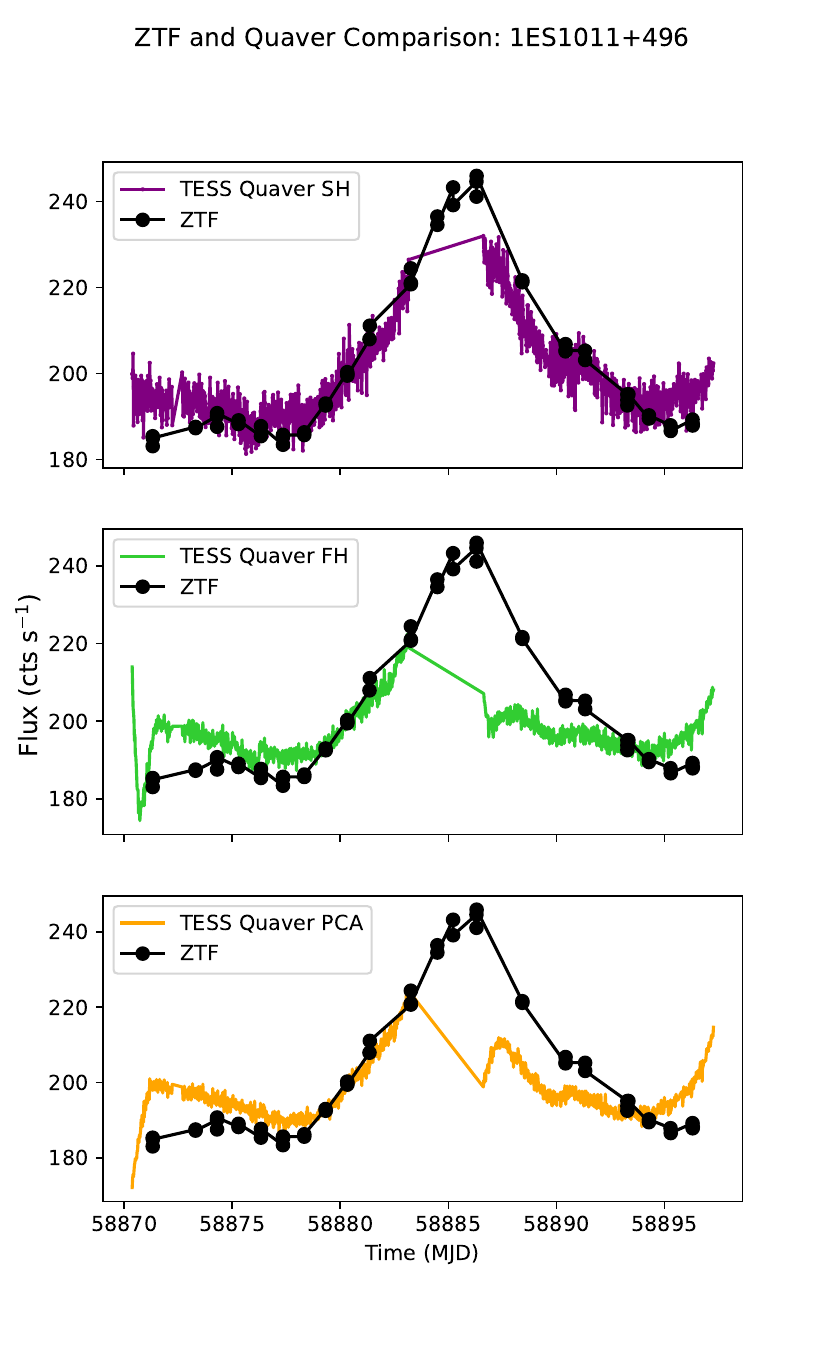} }}%
    \qquad
    \subfloat
    {{\includegraphics[width=8cm]{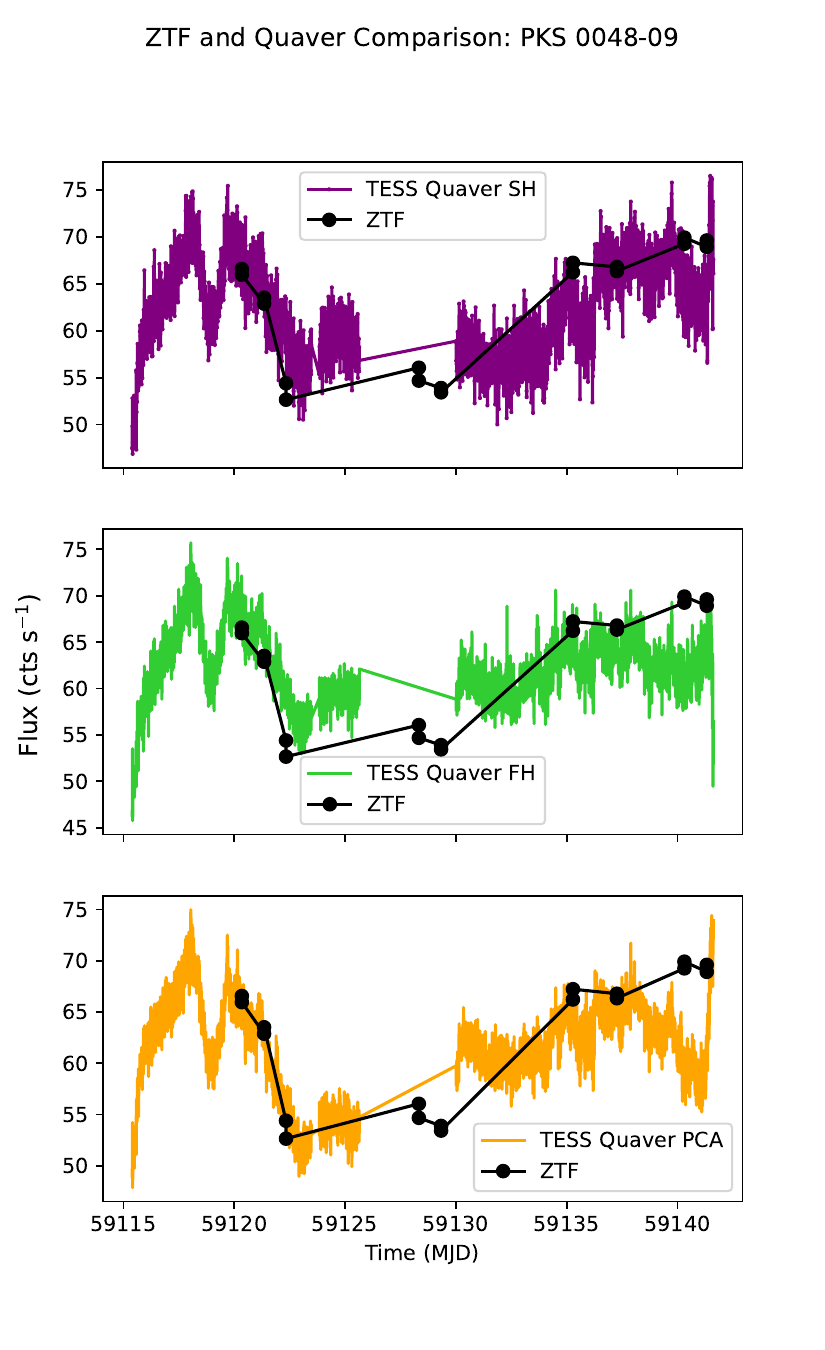} }}%
    \caption{Comparison of the three \texttt{Quaver} reduction methods (colored light curves as indicated in legend) and ground-based light curves from ZTF (black). Light curves have been rescaled to have similar normalizations for comparison of long-term behavior. Conversion of TESS counts to standard magnitude systems is not straightforward due to its unusual bandpass; instead, ZTF magnitudes have been converted to generic flux counts via $f = 10^{m/-2.5}$ and rescaled to a similar normalization as the TESS light curves; see text for details.}%
    \label{fig:quaver_ztf_comp}%
\end{figure*}


The Quaver User Guide available with the code offers a detailed description of each reduction method and a chart explaining their workflows. The best choice of method depends upon the user's science case. In general:

\begin{itemize}
    \item The simple PCA version is the least rigorous but most intuitive and similar to existing methods in the literature. This method is prone to both over-fitting long-term trends as well as missing subtle, typically electronic multiplicative effects. In this method, all pixels (regardless of brightness or faintness) outside the source aperture and its buffer zone are fit by a number of principal components set by the user. These components are stored in a design matrix, which is then used by \texttt{Lightkurve}'s \texttt{RegressionCorrector} class to correct the light curve extracted from the source aperture.

    \item The Simple Hybrid method handles the background effects and instrumental systematics in the ``bright" source-dominated pixels separately, but does not apply a fit to remove the background light from the source itself, instead using simple background subtraction. The light curve of the ``faint" pixels (as defined by the user-adjustable 1.5$\sigma$~ threshold) is \textbf{extracted directly with those pixels as the aperture, scaled multiplicatively to reflect the difference in the number of pixels between the background and source apertures, and then} subtracted from the source light curve. Since no fitting is performed, this is less likely to remove true long-term variability, which can coincidentally mimic the source's true behavior. Once the background is subtracted, the principal components of the faint background pixels are calculated and used to correct the the bright pixels (excluding the source and buffer zone). Then, the principal components of the background-corrected bright pixels are calculated, and stored in a design matrix which is used by \texttt{RegressionCorrector} upon the background-subtracted source light curve, accounting for the multiplicative systematics. Light curves extracted using this method show excellent similarity to simultaneous (but more sparsely-sampled) ground-based light curves.

    \item The Full Hybrid method is the most aggressive. Like the simple hybrid method, it first creates a design matrix housing the principal components of the faint background pixels, uses this matrix to correct the bright pixels, and then derives the principal components of the background-corrected bright pixels and stores them in a design matrix for the multiplicative systematics. However, instead of performing a simple background subtraction, it combines the additive background and multiplicative systematics design matrices into a hybrid \texttt{DesignMatrixCollection} object, which \texttt{RegressionCorrector} then uses to correct the light curve extracted from the source aperture. This method does a very thorough job removing all systematics, but can over-fit long-term behavior in some cases (which is generically true of correction at a higher order than background subtraction). This method is most appropriate for users who wish to study detailed high-frequency behavior, such as flare modeling or searching for rapid quasi-periods, since the subtle effects of electronic cross-talk noise and other high-frequency systematic signals are removed in a sophisticated fashion. This method (along with the PCA method) is risky for users interested in low-frequency power, especially in cases where light curves of multiple sectors will be stitched together.
    
\end{itemize}

Figure~\ref{fig:quaver_ztf_comp} shows the results of each of the three correction methods compared to ground-based light curves of blazars with good simultaneous sampling by the Zwicky Transient Facility \citep[ZTF; ][]{Masci2019}. In order to plot the ZTF and TESS light curves together, some rescaling is required. We convert the ZTF light curves from magnitudes to a generic flux via $F = 10^{m/-2.5}$. \textbf{We then locate datapoints from the TESS light curve that correspond to the ZTF sampling times, and minimize the function $\sum_{t=t_\mathrm{{TESS},i}}^{t_{\mathrm{TESS},f}} (F_\mathrm{TESS} - A * F_\mathrm{ZTF})^2$ to find the best-fitting scale factor $A$ (where $t_\mathrm{{TESS},i}$ and $t_{\mathrm{TESS},f}$ are the first and final timestamps of the TESS monitoring)}. This multiplicative factor is applied to the generic ZTF flux. One can see that the Simple Hybrid method best preserves long-term trends while the Full Hybrid and PCA methods can result in a flatter light curve. Although in many cases these  methods do not overfit (as in PKS~0048-09), other objects experience various degrees of overfitting with methods in which the background is treated using principal component regression (as in 1ES~1011+496, significantly). This overfitting occurs when the principal components have too many degrees of freedom, allowing them to frequently match the object's true variability, which is then removed. Each of the light curves shown in Figure~\ref{fig:quaver_ztf_comp} was fit using three principal components; the user of \texttt{Quaver} is free to use fewer or more depending on their preferences.


\begin{figure*}
\centering
\includegraphics[width=\textwidth]{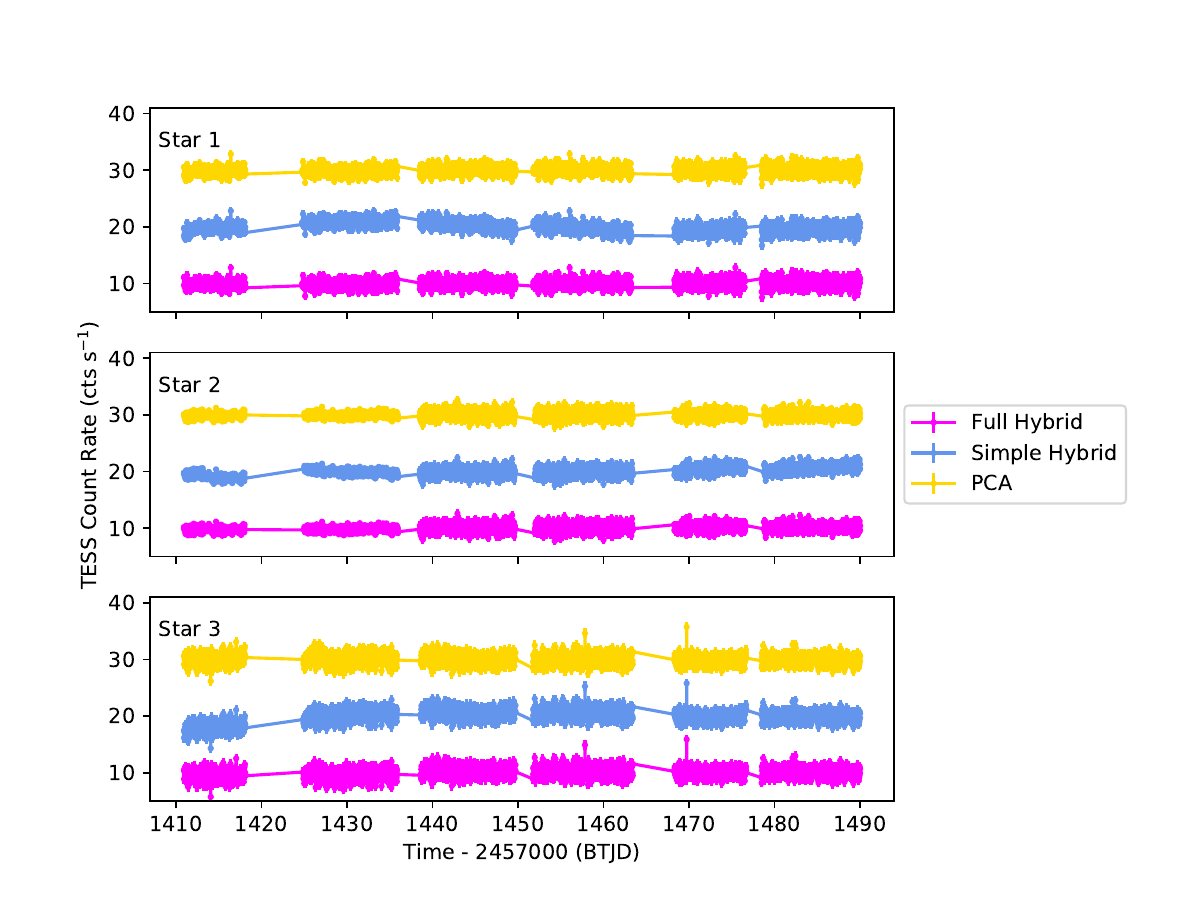}
\caption{Cycle 1 light curves of three non-variable reference stars extracted by the Quaver pipeline. Each panel shows the light curves of stars 1, 2, and 3, as labeled in Figure~\ref{fig:ref_stars_map}, from top to bottom. In each panel, all three Quaver methods are used, with colors given in the legend. Light curves have been arbitrarily offset for visible distinction. Differences in the noise spread of the light curves is due to the different apertures used to extract the stars and variable level of background light from sector to sector.}
\label{fig:ref_stars_lcs}
\end{figure*}

As an indication that the pipeline is capable of removing instrumental and background systematics from targets, we have also extracted the light curves of three stars in the same field of view as Pictor~A, labeled in Figure~\ref{fig:ref_stars_map}. Their light curves for Cycle 1 and Cycle 3 are shown in Figure~\ref{fig:ref_stars_lcs}. Each star was extracted with the same \texttt{Quaver} parameters as Pictor~A. The stellar light curves are flat, as expected for inactive stars, with very few residual effects remaining. Occasional single-cadence spikes remain in all pipeline methods; the Simple~Hybrid method is prone to leaving slight ($<1\%$) ramps across the light curve.

The most dominant term removed by the PCA components is always the scattered light background (see Appendix~\ref{app:scattered_light}). Additional terms are caused by all of the effects listed in Section~\ref{sec:aperture}. Occasionally, a bright nearby source with strong variability may contaminate the target light curve due to overlapping pixel response functions (PRFs). In this case, \texttt{Quaver} is capable of removing the contaminant trend (Appendix~\ref{app:contamination}).

\subsection{Multi-sector Light Curves}
If the light curve spans multiple sectors, then each sector is displayed to the user in turn for custom aperture selection, cadence masking, and extraction. At the end of this process, the sectors are additively stitched together into one light curve, simply by matching the mean of the ten fluxes at either side of the sector gap. Although this method is not ideal, in that it ignores true variability that occurred within the sector gaps, it is the least likely to introduce spurious variability. Tests of more complex methods to model the long-term behavior using, for example, multi-noded splines, resulted in considerable added variability not innate to the source. So, this simplistic stitching remains our current method of choice. Indeed, many authors working on a number of other science applications have also contended with this problem (e.g., \citealt{Claytor2022}, \citealt{Anthony2022}, \citealt{Avallone2022}). In any case, each sector's light curve is also output separately from the full stitched light curve, so the user is free to analyze them separately or stitch them using different methods, as they prefer.

\subsubsection{Assessing the Correction}
\label{correction_diagnostics}
\begin{figure*}
\centering
\includegraphics[width=\textwidth]{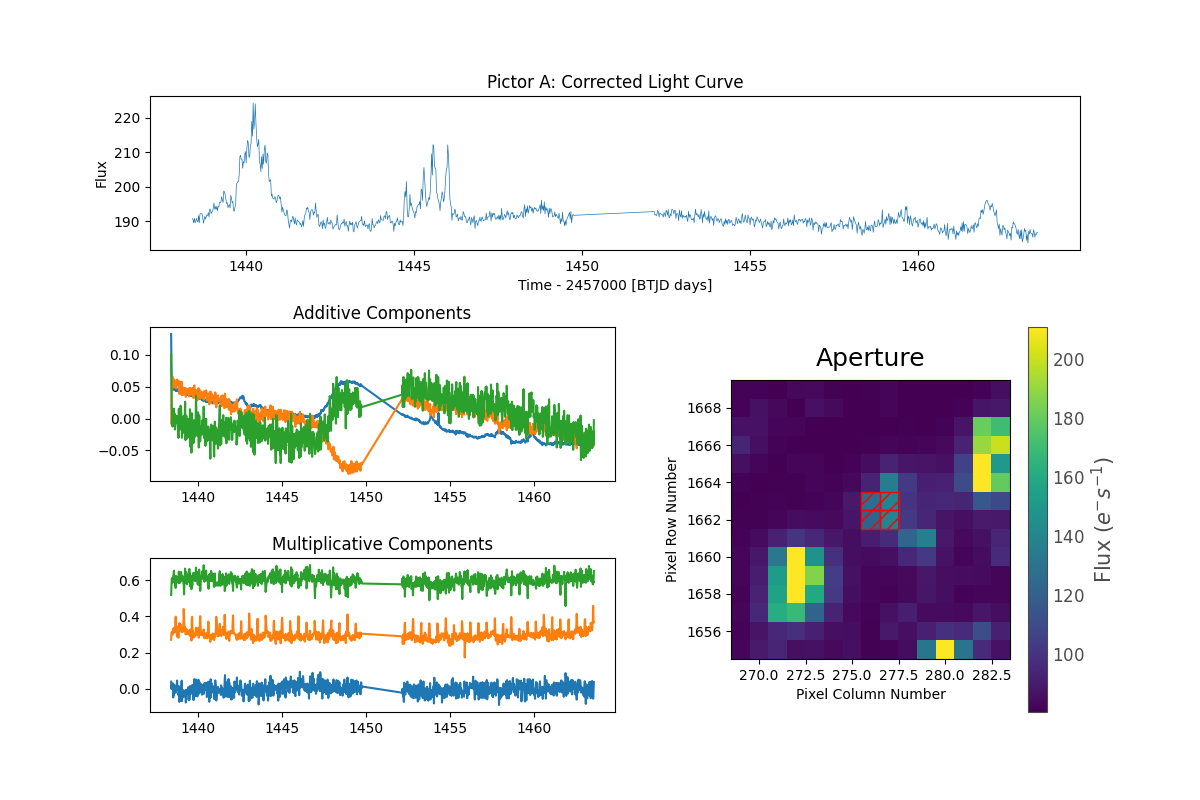}
\caption{Example \texttt{Quaver} output panel for Pictor A in Sector 5. {\it{Top}}: corrected light curve created with the Simple Hybrid method. {\it{Bottom left}}: additive and multiplicative components removed by the hybrid method. Different colors correspond to different PCA components. {\it{Bottom right}}: Postage stamp of the TESS Target Pixel File (TPF). The source is well detected in the center of the image, and the extraction region is cross-hatched in red. The pixel scale is 21 arcsec pixel$^{-1}$.}
\label{fig:tpf}
\end{figure*}

The main output panel for a single-sector \texttt{Quaver} run is shown in Figure~\ref{fig:tpf}. It illustrates the corrected light curve, the additive and multiplicative components identified in the PCA fits, and the chosen aperture.

We include two further outputs from the program to assist users in deciding which method is best for their target, and the overall quality of the correction.

First, in addition to inspecting the correction vectors used in the fit in the standard output panel, it is useful to see the final result of the combined fit that was ultimately removed from the light curve. \texttt{Quaver} produces three output plots, showing the uncorrected light curve, the corrected light curve, and the final form of the fit that was removed (i.e., the linear combination of the constituent vectors with the coefficients that resulted from the fit). Examples for all three methods are shown in Figure~\ref{fig:fitdiag} for Sector~5. Corrections for the other sectors can be found in Appendix~\ref{app:corrections}. We also show these results for two of the stars and Sectors from Figure~\ref{fig:ref_stars_lcs}.

All three pipeline versions remove similar long-term trends; the background light curve subtracted by the Simple Hybrid method is the dominant term in the total fit in the other methods, as expected. One can also see that the variable star that dominates the second multiplicative correction vector (Figure~\ref{fig:tpf}) has not received any significant weighting in the final fit (it cannot be seen in the multiplicative model); therefore, this signal was not injected into the source light curve.

The second output useful for assessing the fit is the overfitting metric defined in \texttt{Lightkurve}\footnote{\url{https://docs.lightkurve.org/tutorials/2-creating-light-curves/2-3-how-to-use-cbvcorrector.html}}, which compares the broadband power in a Lomb-Scargle periodogram before and after correction. This metric value is output to the terminal after each Sector is completed, for each reduction method. Higher values are better; a metric of 0.5 indicates that power from injected noise due to fitting is equal to power from light curve uncertainties. The documentation for LightKurve indicates that values above 0.8 mean that no significant over-fitting has occurred. The user can compare the overfitting metric of the three \texttt{Quaver} methods in determining which to use, or simply to assess the amount of overfitting present in their chosen output. We present the results of these metric tests in Table~\ref{t:fitmetrics}.

\begin{table}[t]
\centering 
\begin{tabular}{c c c c} 
\hline\hline 
Sector & Simple PCA & Simple Hybrid & Full Hybrid \\ [0.5ex] 
\hline 
4 & 0.73 & 0.86 & 0.74 \\ 
5 & 0.65 & 0.88 & 0.52 \\
6 & 0.82 & 0.91 & 0.79 \\
31 & 0.84 & 0.81 & 0.83 \\
33 & 0.80 & 0.81 & 0.70 \\ [1ex] 
\hline 
\end{tabular}
\caption{Overfitting metrics obtained using the comparison of broadband power from Lomb-Scargle periodograms. Higher numbers indicate better fits with a lower likelihood of over-fitting; numbers over 0.5 are acceptable, while numbers over 0.8 are good.}
\label{t:fitmetrics} 
\end{table}

\begin{figure*}[htp]
\centering
\subfloat{%
  \includegraphics[width=0.8\textwidth]{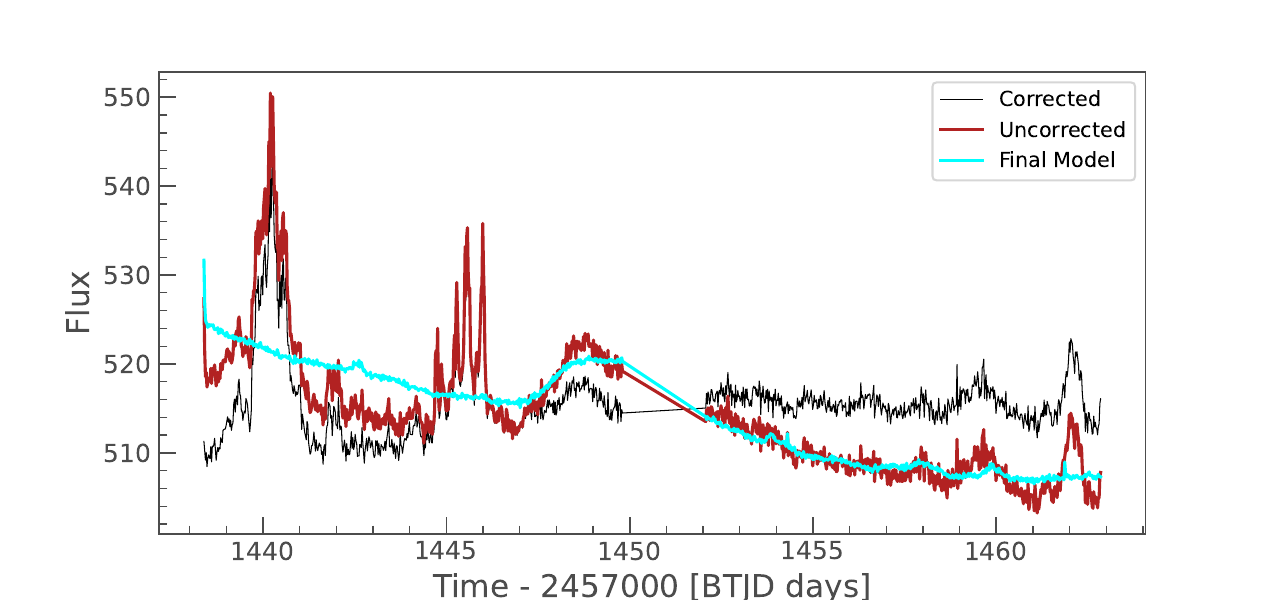}%
}

\subfloat{%
  \includegraphics[width=0.8\textwidth]{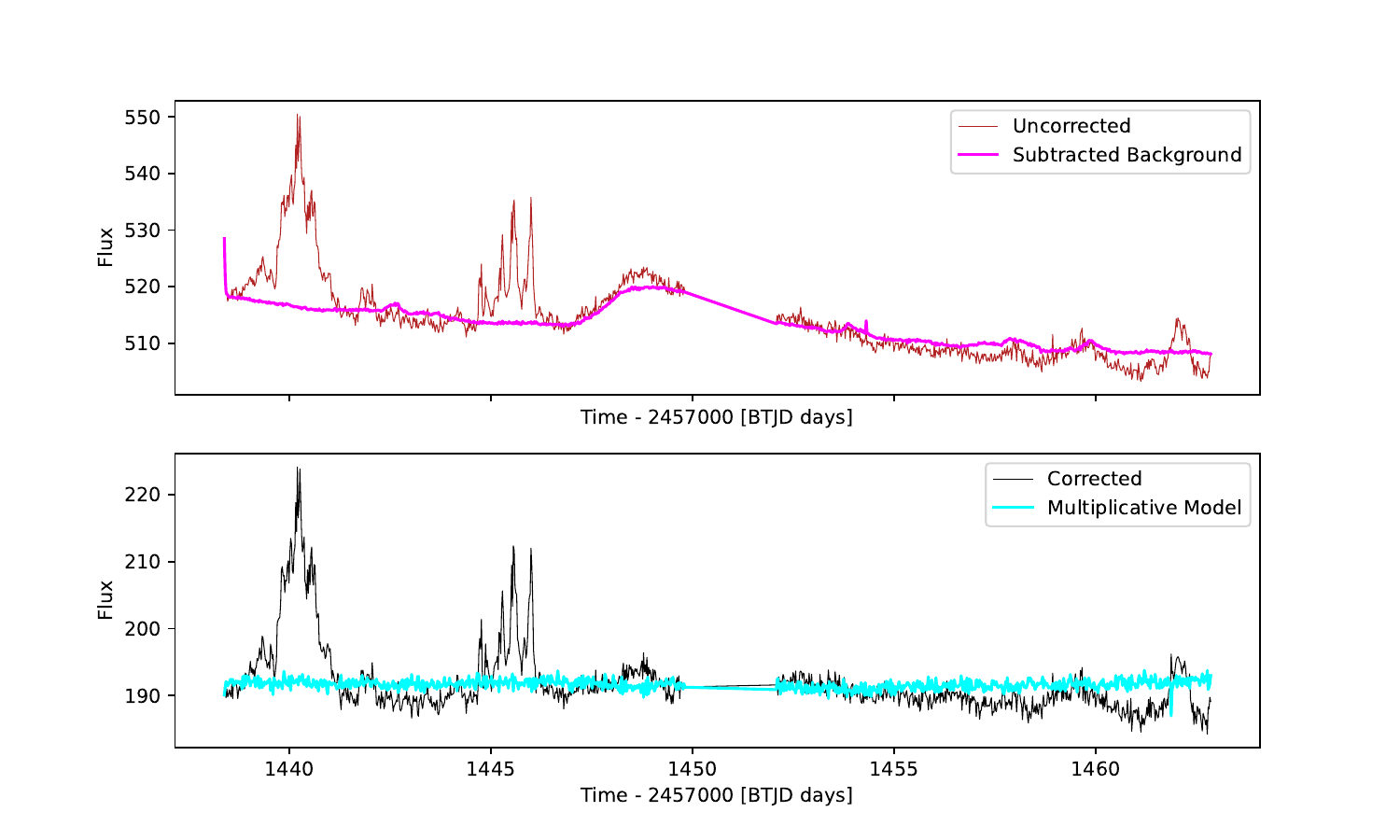}%
}

\subfloat{%
  \includegraphics[width=0.8\textwidth]{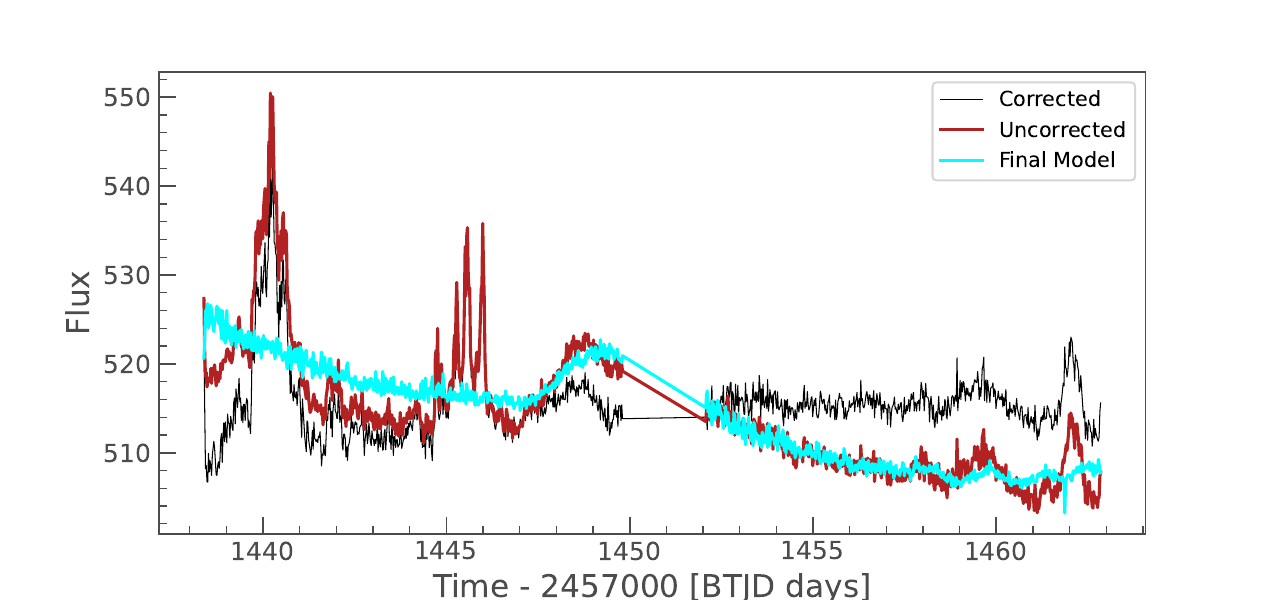}%
}

\caption{Fit output diagrams for Sector~5 for each of the three methods. \emph{Top}: PCA fitting only. \emph{Middle}: Simple hybrid model, with subtracted background light curve \textbf{(scaled to match the median flux of the raw light curve)} shown with the original in the top panel, and the background-subtracted light curve shown with the corrected multiplicative model in the lower panel. \textbf{The amplitude of the subtracted background can be estimated by comparing the raw and background-subtracted curves.} \emph{Bottom}: Full hybrid model.}
\label{fig:fitdiag}
\end{figure*}


\section{Cautions and Tips}
\label{sec:cautions}

\subsection{Spacecraft systematics}
We caution that \texttt{Quaver} is not able to completely remove and identify all spacecraft systematics. To the extent that pointing jitter, velocity aberration, and thermal breathing affect all sources in the field in a similar way, \texttt{Quaver} can identify and remove these trends. However, because AGN hosts are typically not point sources (while most of the field sources used to determine the correction vectors are likely to be stars), and because individual pixels are not identical in their responses across the detector, traces of these effects can remain, especially in sources where a small extraction aperture was required.

\subsection{Underfitting}
The reduction of any light curve involves striking a balance between over- and under-fitting, an especially challenging problem to diagnose when your target varies stochastically and without quiet periods or predictive behavior. While using fewer PCA vectors can reduce the chances of over-fitting, using too few can result in in under-fitting, in which systematics remain in the light curve. Of \texttt{Quaver}'s methods, the Simple Hybrid approach is the most prone to under-fitting long-term trends, as a simple background subtraction is more likely to leave residual trends in the light curve than a fitted approach.

There is an underfitting metric provided by \texttt{Lightkurve}, alongside the over-fitting metric mentioned in the previous section. However, it compares the output light curve to data reduced by the TESS SPOC, and warns that it is in appropriate to compare these to light curves reduced in other ways. These targets are also almost certainly much brighter than the majority of AGN, so we do not include this metric in \texttt{Quaver} at present. Future versions may include a custom underfitting metric designed for \texttt{Quaver} output.

\subsection{Signal Injection Using PCA}
In some rare cases, if a strongly variable source in the field is the most dominant component in the regression matrix, such signal could theoretically be \emph{injected} into the light curve during the correction. It is important for the user to inspect the correction vectors in the output image (e.g., Figure~\ref{fig:tpf}), as well as the form of the final fitted model that will be removed from the light curve. It is occiasionally the case that a variable target will manifest in the multiplicative components, as in the orange vector in the bottom panel of Figure~\ref{fig:tpf}. However, when the set of vectors is regressed against the light curve, if these signals are not present, those vectors are down-weighted severely. This can be seen by inspecting the final models shown in Figure~\ref{fig:fitdiag}, which do not include this periodic component.

\subsection{Effects of Scientific Analyses of AGN}

The purpose of this pipeline is to create a user-friendly tool that allows for flexibility and transparency in dealing with TESS data. It is not intended to automatically create trustworthy light curves using the default parameters. 

The process therefore includes a number of subjective steps (aperture selection, cadence masking, number of principal components, type of reduction method, etc), which are often efficiently decided automatically using pre-defined criteria in other TESS pipelines \citep[for example, \texttt{eleanor; }][]{Feinstein2019}. When making use of the additional flexibility of Quaver, the user must think carefully about the effects of each reduction choice on the desired scientific result.

Using too many principal components is likely to result in over-fitting, especially of low-frequency behavior from long-term trends. This has a major effect on the shape of the power spectrum at low frequencies, and can even induce a break timescale where one did not exist before. Because these break timescales are often the goal of AGN timing, as they potentially offer an independent measure of the black hole mass or accretion behavior, it is extremely important to avoid over-fitting gradual trends as much as possible. This is the motivation behind our choice of the Simple Hybrid method for this work, as it is the least likely to over-fit at low frequencies. 

Underfitting also affects the shape of the power spectrum, by leaving power at many frequencies that should have been removed. \texttt{Quaver} is effective at removing high-frequency behavior even in the Simple Hybrid method, so underfitting at high frequencies is unlikely, but very short-term spikes or jitter can contribute to high frequency power and artificially shallow the slope of the PSD. 

Finally, we stress that TESS is an instrument of relative photometry. In cases like AGN, when the a-priori behavior is difficult or impossible to model and the ground-truth may not be known, and especially when detrending methods and subjective aperture sizes can affect the baseline flux, it is important to avoid using the data for science cases requiring \emph{absolute} photometry, such as estimating intrinsic AGN luminosities. In these cases, it is important to use ground-based data like ZTF to normalize the light curve \textbf{accurately} before proceeding.

\begin{figure*}[htp]

\subfloat{%
  \includegraphics[width=\textwidth]{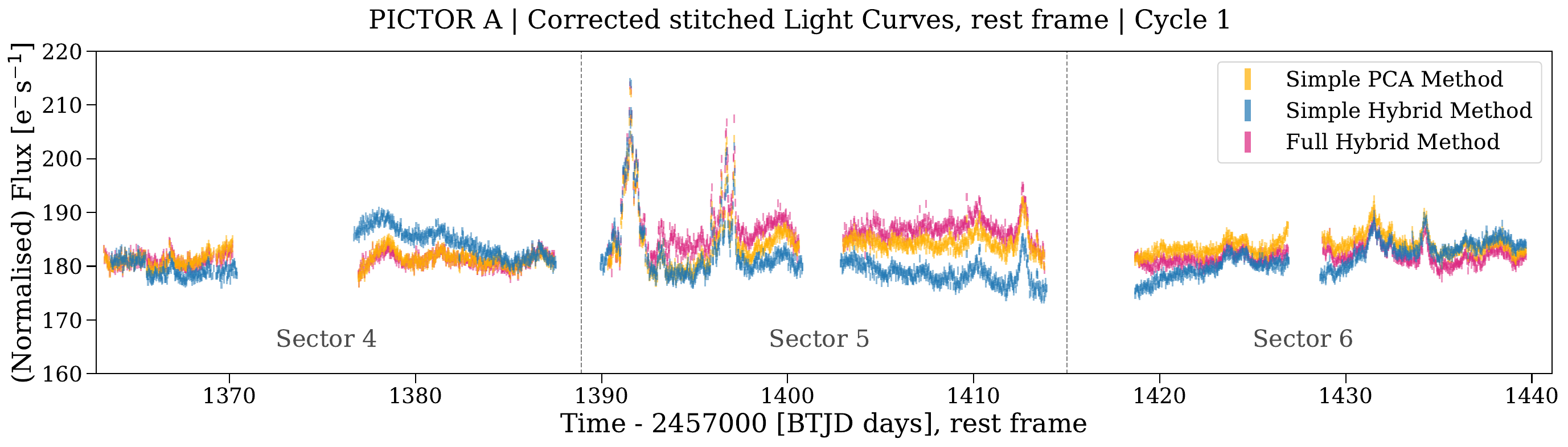}%
}

\subfloat{%
  \includegraphics[width=\textwidth]{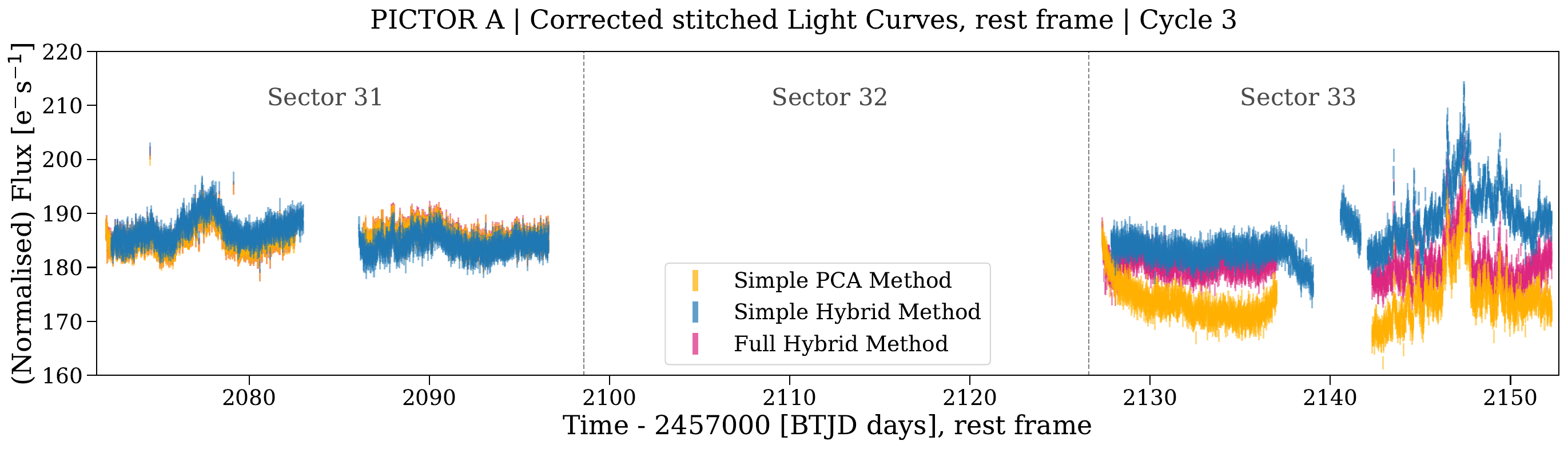}%
}

\caption{Stitched light curves of Pictor A in Cycle 1 (Sectors 4, 5, 6; top panel) and Cycle 3 (Sectors 31 and 33; bottom panel). The time axis has been shifted to the galaxy's rest frame. The colors correspond to the different extraction methods: Simple PCA (orange), Simple Hybrid (blue) and Full Hybrid (purple). In addition to the normalisation applied for the stitching (see text for details), the Simple PCA and Full Hybrid light curves are artificially normalised by subtracting a factor of $\sim$ 285 to match the Simply Hybrid light curve. We notice that while the main features (e.g. flares) are similar in each light curve, the Simple Hybrid one shows more variability at timescales comparable to the sector's length. For the following analysis we will use the light curve extracted with the Simple Hybrid method.}
\label{fig:pictora_lcs}
\end{figure*}

\subsection{Light Curve Errors}
\label{sec:lc_errors}

\textbf{The photometric errors on the raw extracted light curve are propagated for each cadence from the errors of the flux in each pixel. The errors given in the final, corrected light curve can either be propagated through the reduction and correction process, or can simply reproduce the original photometric errors from the raw light curve.} The tunable parameter \texttt{prop\_error\_flag} allows the user to decide whether to allow \texttt{Lightkurve} to propagate the errors through the matrix regression process, or to retain the original errors. Propagation increases the program runtime.

For the purposes of this paper, the original photometric errors are maintained, as this is the typical \texttt{Lightkurve} use case. The magnitude of the errors does not change by more than 0.2~dex, in the case of Pictor~A, when the errors are propagated.

\section{Variability measurements}\label{sec:vm}

The light curves of Pictor~A as extracted with \texttt{Quaver} with all three methods are shown in Figure \ref{fig:pictora_lcs}. We extracted these light curves using \texttt{Quaver}'s default fitting parameters and our interactively-selected 2x2 pixel extraction aperture (Fig. \ref{fig:tpf}). 
Our present science application is a low-mass Seyfert galaxy, with expected variations on the timescales of days to weeks. Based on the reasoning in the preceding section, we have selected the Simple Hybrid method. Table~\ref{t:fitmetrics} indicates that the Simple Hybrid method is least likely to overfit this target, as it is the only method with metrics above 0.8 in all sectors. We are interested in not only the high-frequency slope of the power spectrum, which is well-preserved in all methods, but also whether the power spectrum turns over towards low frequencies at some characteristic timescale. The risk of over-fitting low-frequency trends that is evident in the Full Hybrid method and removing intrinsic low-frequency power is not appropriate for such an investigation. Although the Simple Hybrid method can be prone to under-fitting these trends, inspection of the three different methods in Figure~\ref{fig:fitdiag} indicates that the Simple Hybrid method does a good job of removing the same low-frequency behavior as the simple PCA method.

In the following sections we describe different analyses we performed in order to characterise the variability we observe in the TESS light curves of Pictor~A. Because of the uncertainties arising when trying to stitch together the light curves obtained for different sectors, we  consider each sector separately, therefore concentrating on timescales up to 26 days (in the galaxy's rest frame). 

\subsection{Flux distributions and the RMS-flux relation}

\begin{figure*}
\centering
\includegraphics[width=\textwidth]{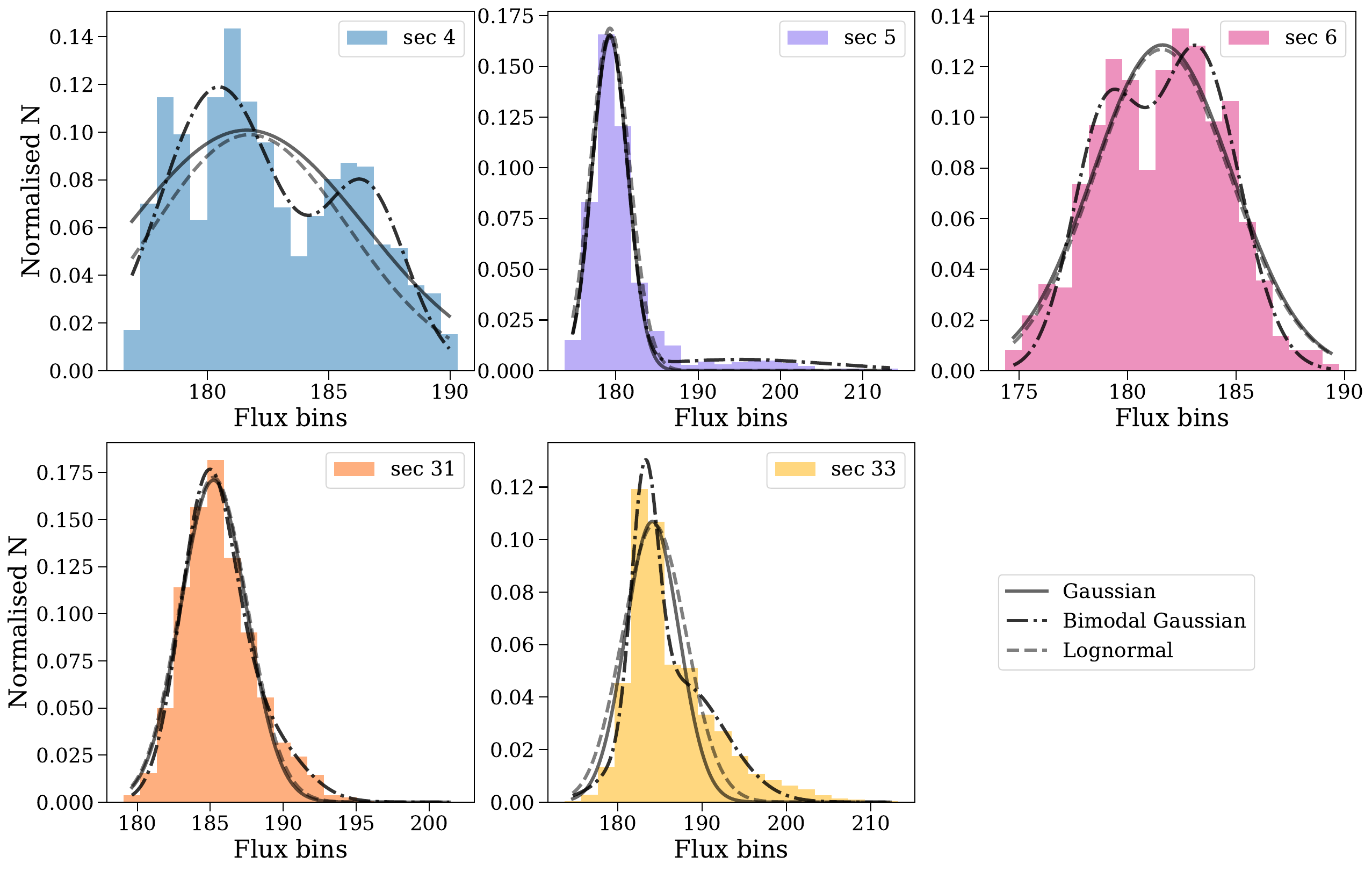}
\caption{Flux distribution histograms in the individual sectors with Gaussian (solid line), bimodal Gaussian (dash-dotted line) and Lognormal (dashed line) fits.}
\label{fig:fluxdist}
\end{figure*}

Fig. \ref{fig:fluxdist} shows the flux distributions for the individual sectors, which we attempted to fit with Gaussian, bimodal and log-normal distributions, as these have all been observed for AGN light curves obtained at different wavelengths (e.g. \citealt{Uttley2005}, \citealt{MacLeod2010}, \citealt{Smith2018}, \citealt{BhattaDithal2020}). Log-normal distributions are indeed predicted by multiplicative fluctuation models, but are \emph{not} themselves necessarily indicative of such underlying processes \citep{Scargle2020}. Gaussian and bimodal distributions may be observed in the case that multiple components are contributing to the observed variability. While sectors 5, 31 and 33 are well fit with a bimodal distribution, the flux distribution in sectors 4 and 6 is less well defined. This may arise from the fact that in these sectors more cadences had to be excluded because of systematic effects (see Section \ref{sec:quaver}), and the number of observations is therefore too low to confidently probe the underlying flux distribution. 

Similarly to what was observed in {\it{Kepler}} AGN light curves (e.g., \citealt{Smith2018}), our data do not show any evidence of the rms-flux relation exhibited by the X-ray light curves of AGN (e.g., \citealt{Vaughan2003}, \citealt{McHardy2004}) and X-ray binaries (e.g., \citealt{Uttley2001}).

\subsection{Periodograms}\label{sec:per}

\begin{figure*}
\centering
\includegraphics[width=\textwidth]{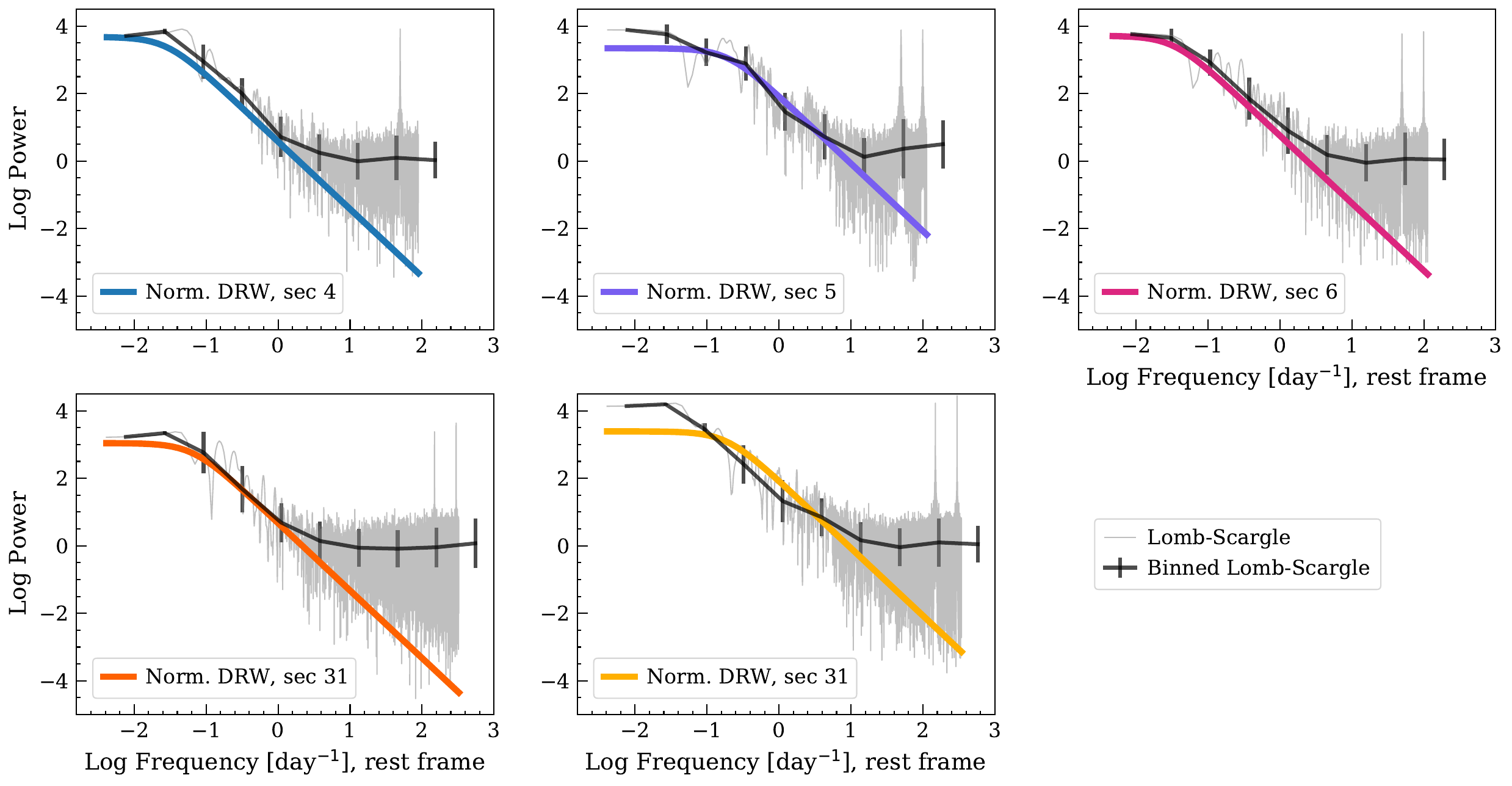}
\caption{Lomb-Scargle periodograms computed for the sectors separately. Gray shadow: total computed periodograms. The peaks at the highest frequencies are aliases due to the time resolution of the TESS observations (30 minutes in Cycle 1 and 10 minutes in Cycle 3). Black line: binned Lomb-Scargle periodograms. \textbf{Here, the error bars represent the scatter within each bin.} Coloured lines (same colour coding as Fig. \ref{fig:fluxdist}): normalised periodogram for the underlying DRW model. \textbf{These models are more reliable than the Lomb-Scargle estimates (see Section \ref{sec:per} for more details).}}
\label{fig:per_1}
\end{figure*}

AGN periodograms (or power spectra) are usually well fit by either a single or broken power law, where the break timescale may be associated to the characteristic timescales of the underlying variability process. Since the extracted TESS light curves are not evenly spaced (because of the observing gaps and excluded cadences), we computed the periodogram of each sector following the generalised Lomb-Scargle definition (\citealt{Cumming2008} with formalism from \citealt{Zechmeister2009}) which is less sensitive to gaps in the light curve compared to standard Fourier techniques. In Fig. \ref{fig:per_1}, the measured Lomb-Scargle periodogram of each sector is shown in gray, while the black lines show the corresponding binned periodograms. As we will better quantify below, each periodogram shows a bent power-law behavior at low frequencies and flattens at higher frequencies due to Poisson noise.

\begin{figure*}
\centering
\includegraphics[width=\textwidth]{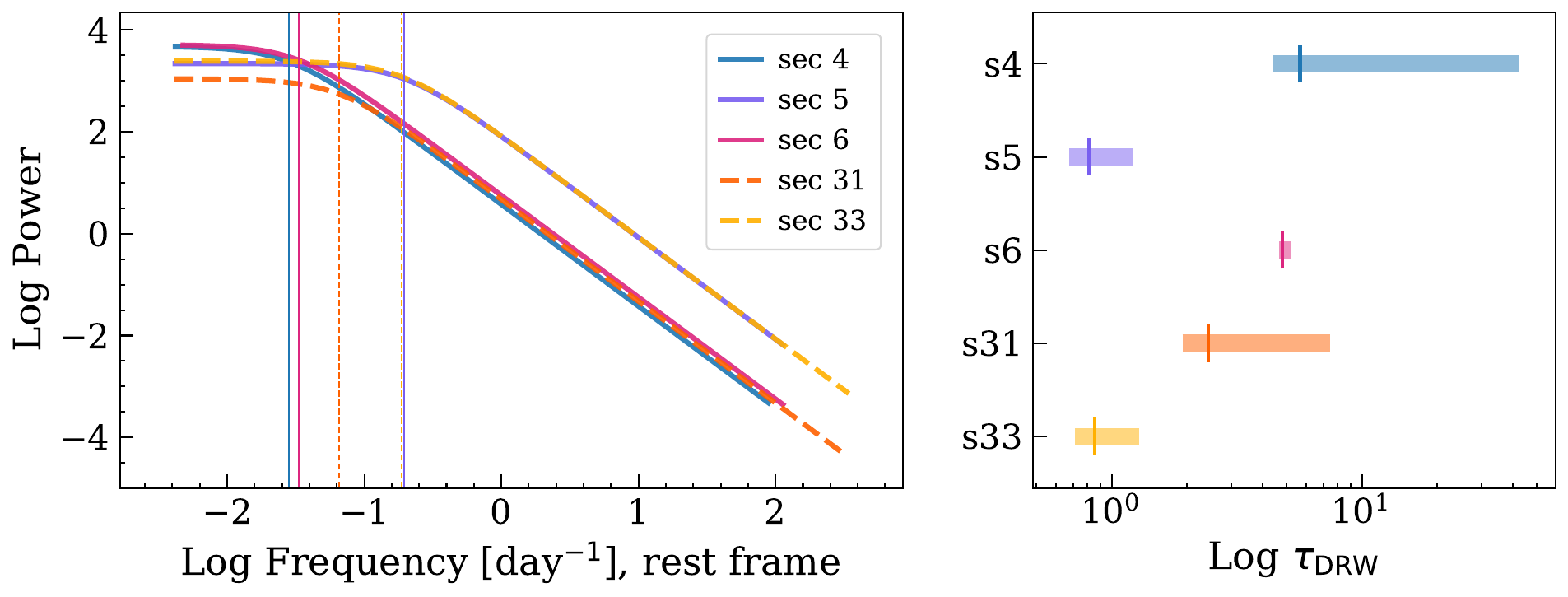}
\caption{{\it{Left}: } Periodograms from the DRW fit of the different sectors. {\it{Right}: } Corresponding $\tau_{\rm {DRW}}$. Depending on the sector, $\tau_{\rm {DRW}}$ ranges between 0.8 and 5.5 days.}
\label{fig:per_4}
\end{figure*}

Modelling the power spectrum of short light curves in the frequency domain is complicated by spectral distortions such as red noise leak (\citealt{Deeter1982}, \citealt{Deeter1984}) and aliasing (\citealt{Kirchner2005}). We therefore prefer to model the light curve directly in the time domain assuming a damped random walk (DRW\footnote{We note that the DRW model corresponds to the first-order continuous autoregressive model, CARMA(1,0), for irregularly spaced observations.}, \citealt{Kelly2014}) model using the \texttt{EzTao} Python package (\citealt{Yu2022}), and to compute the corresponding power spectrum, which we then compare to the Lomb-Scargle periodogram as a sanity check, as done in, for example,  \cite{Burke2020}. An example of DRW fit results, including residuals and posteriors distributions, is showed in Appendix \ref{app:DRW}. We further tested that the DRW is a good model for our data by applying a Durbin-Watson test (\citealt{Durbin1950, Durbin1951}) to verify that the residuals are not autocorrelated. The Durbin-Watson test returns values between $d = 2$ and $d = 2.5$ for 4 out of 5 of our TESS Sectors, and $d = 2.8$ for the remaining Sector 5. As a rule of thumb, $d = 2$ (or close to this value) indicates no autocorrelation, while $d > 3$ ($d <1$) may be a cause of alarm for negative (positive) autocorrelation. Since the values we obtain for our data are in the range $2 < d < 3$, we are confident that the residuals are not (strongly) autocorrelated. This, together with the fact that the residuals are normally distributed, supports our hypothesis that the DRW is an acceptable model for our data. The obtained periodograms are shown in Fig.~\ref{fig:per_1} and Fig.~\ref{fig:per_4}, along with the corresponding characteristic timescales $\tau_{\rm DRW}$. The DRW slope in the regime which is not dominated by Poisson noise and/or spectral distortions is consistent with the slopes of the Lomb-Scargle diagram in all the sectors. In addition, we observe that the $\tau_{\rm DRW}$ values are different for sectors 5 and 33, in which some flares are prominent, and sectors 4, 6, and 31, which are mostly devoid of flares: $\tau_{\rm DRW} \sim 0.8$ days for 5 and 33, a $\tau_{\rm DRW} \sim 3-6$ days for the others. This may signify that the AGN is switching between two different variability modes, as we will discuss in Section \ref{sec:disc}. It is important to notice that $\tau_{\rm DRW}$ can usually be fully trusted only if it is shorter than $10\%$ of the considered light curve length $L_{lc}$, in the case of our observations 2.6 days (rest frame). Moreover, if the underlying DRW process has a $\tau_{\rm DRW}$ longer than $L_{lc}$, then the fit will return a value which is $20-30\%$ of $L_{lc}$ (e.g., \citealt{Kozlowski2010}, \citealt{Emmanoulopoulos2013}, \citealt{Burke2020}). The characteristic timescales we measure lie in this critical regime, and should therefore be treated carefully. To check the robustness of our results, for each sector we first simulated 5000 light curves with the same time steps and DRW parameters as the real observations, and then fit them in the same way as done for our data. For each sector, the characteristic timescale $\tau_{\rm DRW}$ measured from the real data and from the simulated ones correspond well within $1\sigma$. As a complementary test,  we run simulations to determine the probability of obtaining the measured $\tau_{\rm DRW}$ in the case that the real $\tau_{\rm DRW}$ is actually longer than the observed time baseline. Specifically, for each sector we simulated three sets of 1000 DRW time series each with same time baseline as our observations and $\tau_{\rm DRW} = 10, 50$ and 100 days, respectively, and fit them following the same procedure as for the real data. For each simulation set we then determined the fraction of simulations for which the fit returns $\tau_{\rm DRW}$ values within a given percent of the $\tau_{\rm DRW}$ measured from the real observations. In summary, we see that for $\tau_{\rm DRW} = 10$ days, less than $10\%$ ($20\%$) of the simulated $\tau_{\rm DRW}$ are within $\pm  10\%$ ($\pm  20\%$) of the $\tau_{\rm DRW}$ measured from real data for the considered period. These percentages strongly decrease for $\tau_{\rm DRW} = 50-100$ days\footnote{Example: If the measured $\tau_{\rm DRW}$ for the considered sector is $\sim5$ days, then less $<10\%$ of the simulations will return a $\tau_{\rm DRW}$ within $5 \pm 0.5$ days, i.e 4.5 to 5.5 days.}. Although the performed tests do not allow us to fully rule out the possibility that the measured $\tau_{\rm DRW}$ are due to biases in the modeling, our results are consistent with the idea that the measured $\tau_{\rm DRW}$ are real.  This is further supported by the agreement between the break frequency observed in the Lomb-Scargle and in the DRW periodograms (Fig. \ref{fig:per_1}). In addition, as we will elaborate in Section \ref{sec:disc}, the measured $\tau_{\rm DRW}$ are consistent with the orbital and thermal timescales computed for Pictor~A, and are therefore  physically meaningful. Finally, we tested that the characteristic timescales measured for the sectors that exhibit strong flares (5, 33) are not driven by the flares themselves. For this purpose, we repeated the DRW fit of sectors 5 and 33 after masking the regions with the most prominent flares. The masking was achieved by fitting the five highest-amplitude flares in these sectors with a typical flare profile:

  \[ F(t) = F_c + \frac{2F_\mathrm{max}}{e^{(t_0 - t)/T_r} + e^{(t-t_0)/T_d}}\]

where $F_c$ is a constant baseline, $F_\mathrm{max}$ is the peak flux of the flare, $t_0$ is the peak time of the flare, and $T_r$ and $T_d$ are the rise and decay times. This asymmetric exponential form was originally used to model gamma ray flares of blazars in $Fermi$ data \citep{Abdo2010}, and was  used to model the flares of the only monitored blazar in the \emph{Kepler} field, W2R~1926+42, by \citet{Li2018}. Once the flares were modeled, we simply removed all data points within the span of the flare's modeled baseline and repeated the DRW fit.

The obtained timescales, $\tau_{\rm DRW} \sim 1-1.2$ days, are consistent within $1\sigma$ with the values measured for these sectors before flare-masking ($\tau_{\rm DRW} \sim 0.8$ days) and well below the values measured for the sectors without strong flares ($\tau_{\rm DRW} \sim 3-6$ days). We also applied the Durbin-Watson tests to the residuals in these cases and obtained $d = 2.35$ for both. These results show that the DRW fit is not driven by the presence of flares, and are consistent with the possibility that the flares may indeed occur during a different variability regime.

\section{Discussion}\label{sec:disc}


As presented in Section \ref{sec:vm}, the flux distributions obtained for the single sectors can mostly be fit by bimodal distributions, instead of simple Gaussian or log-normal functions. In addition, we do not find any evidence of the rms-flux relation observed in X-ray light curves (e.g., \citealt{Uttley2001}, \citealt{Vaughan2003}, \citealt{McHardy2004}). This may indicate that the optical/UV variability observed by {\it{TESS}} on day-month timescales is due to multiple physical processes. In addition, as found in \cite{Smith2018}, the lack of correlation between flux and rms in the {\it{Kepler}} and {\it{TESS}} light curves may indicate that the optical/UV variability is not a simple reprocessing of the more rapid variability observed in the X-ray.

As shown in Fig. \ref{fig:per_4}, the DRW periodograms obtained by fitting the sectors with prominent flares have shorter characteristic timescales ($\tau_{\rm DRW} \sim 0.8$ days) than the sectors without obvious flares ($\tau_{\rm DRW} \sim 3-6$ days). This result persists even when the flares are removed from the light curve, indicating that the flares themselves are not responsible for the shorter timescales. While the TESS sector boundaries are of course artificial, as we are examining the sectors separately, they provide a simple, consistent baseline for comparison between flaring and non-flaring periods, which is otherwise arbitrary. This may signify that the AGN is switching between two different variability modes, as it is often observed in X-ray binaries (e.g., \citealt{Coriat2011}, \citealt{Ecksall2015}), and is expected to happen also in AGN, although at different timescales (e.g., \citealt{Alexander2012}, \citealt{Sartori2016}).

\subsection{Characteristic Timescales}

To put the observed characteristic timescales into context, it is interesting to investigate the different variability timescales expected for Pictor~A given its black hole mass $M_{\rm BH}~=~5.9~\times~10^6 M_{\odot}$ \citep{Koss2022a}. The {\it{light-crossing timescale}} $t_{\textrm{\footnotesize{lc}}}$ is defined as the time needed for light to cross a region of radius $r$, and is therefore the relevant timescale for irradiated discs (e.g., \citealt{Gaskell2003}). For a standard accretion disc (\citealt{Shakura1973}), this is parametrised as:


\[ t_{\textrm{\footnotesize{lc}}} \equiv \frac{r}{c}\simeq 0.875 \left( \frac{r}{150 r_\textrm{\footnotesize{g}}}   \right) M_{\textrm{\footnotesize{8}}} \mbox{           [d]}  \]

where $c$ is the speed of light, $M_{\textrm{\footnotesize{8}}}$ the BH mass in units of $10^8M_{\odot}$, and $r_\textrm{\footnotesize{g}}$ is the gravitational radius, which for the SMBH in Pictor~A corresponds to $r_\textrm{\footnotesize{g}} = 8.7 \times 10^9$m. For a near-UV-to-optical continuum emitting region with $r = 150 r_\textrm{\footnotesize{g}}$ we obtain $t_{\textrm{\footnotesize{lc}}}~=~0.05$~days, which is much shorter than the observed timescales and within the Poisson noise for these light curves in any case.

\noindent
Another important timescale is the {\it{dynamical timescale}} $t_{\textrm{\footnotesize{dyn}}}$, which is the time needed for the accretion disc to achieve hydrostatic equilibrium:


\[ t_{\textrm{\footnotesize{dyn}}} \equiv \frac{r}{v} \simeq \frac{1}{\Omega} \simeq 10 \left( \frac{r}{150 r_\textrm{\footnotesize{g}}} \right)^{3/2}M_{\textrm{\footnotesize{8}}} \mbox{           [d]}  \]

where $\Omega = (G M_{\rm BH} / r^3)^{1/2}$ is the angular velocity of the material within the disc (\citealt{Netzer2013}) and $G$ the gravitational constant. This is also considered to be the shortest timescale on which it is possible to observe large-scale physical changes in the disc. For the near-UV-to-optical continuum emitting region with $r = 150 r_\textrm{\footnotesize{g}}$ we obtain $t_{\textrm{\footnotesize{dyn}}} = 0.6$ days.

Closely related to the dynamical timescale is the {\it{orbital timescale}}, defined as:

\[t_{\textrm{\footnotesize{orb}}} = 2 \pi t_{\textrm{\footnotesize{dyn}}}\]

For the same near-UV-to-optical continuum emitting region as above we obtain $t_{\textrm{\footnotesize{orb}}} = 3.7$ days.

Finally, the {\it{thermal timescale}} $t_{\textrm{\footnotesize{therm}}}$ is the time needed for energy to redistribute due to
dissipative and cooling processes within the disc, and it is related to $t_{\textrm{\footnotesize{dyn}}}$ and the viscosity $\alpha$ as:

\[ t_{\textrm{\footnotesize{therm}}} = \frac{t_{\textrm{\footnotesize{dyn}}}}{\alpha}\]

Assuming $\alpha = 0.1 - 0.4$ (\citealt{King2007}) and the dynamical timescale computed above  we obtain $t_{\textrm{\footnotesize{therm}}} = 1.5 - 5.9$ days.

It is important to recognize that the characteristic timescale $\tau_\mathrm{DRW}$ simply parameterizes the phenomenological damped random walk model to fit the data, in a statistical sense. While this may generate a range for a physical ``characteristic timescale" of variability, it is \emph{not} predicted by any specific physical models. We can therefore only determine whether the $\tau_\mathrm{DRW}$ is plausibly consistent with physically relevant timescales; it cannot be used to eliminate or support different physical origin models of the variability. The purpose of our analysis is to demonstrate the potential and limitations of TESS data in studying these phenomena. The characteristic timescales resulting from our DRW analysis of Pictor~A are consistent with the dynamical, orbital and thermal timescales for the best-available black hole mass estimate, as found in numerous previous studies of AGN variability. Indeed, recent studies pointed out that the variability observed in the optical AGN spectrum, both as regards stochastic variability and changing-look (of changing-state) AGN, is mostly consistent with the thermal timescale (e.g., \citealt{Kelly2009}, \citealt{Stern2018}, \citealt{Parker2019}). This would imply that variability may be mostly due to
rapid temperature changes in large parts of the accretion disc (e.g., \textbf{\citealt{Dexter2011}}, \citealt{Ricci2022}).

\section{Summary and Conclusions}
\label{sec:conclusion}

In this paper we present \texttt{Quaver}, a new software tool designed specifically to extract {\it{TESS}} light curves of extended and faint sources exhibiting stochastic variability, as is the case for AGN. The code is designed to be fully interactive and flexible, allowing the user to chose the best extraction parameters for their data, and is publicly available on Github\footnote{\url{https://github.com/kristalynnesmith/quaver}}.

As a first example, we used \texttt{Quaver} to extract the {\it{TESS}} light curves available for the radio-loud AGN galaxy Pictor~A, and perform a variability analysis. The obtained light curves are well fit with a DRW model. We find that the source exhibits characteristic timescales $\tau_{\rm DRW} \sim 3-6$ days during periods when flares are not dominant, and   $\tau_{\rm DRW} \sim 1$ day during periods with numerous flares, even when the flares are removed from the DRW analysis. The observed timescales are consistent with the dynamical, orbital and thermal timescales expected for the BH mass of Pictor~A, as observed also in other sources by recent optical/UV variability studies. 

This work demonstrates the power of \texttt{Quaver} for extracting {\it{TESS}} light curves and therefore enabling accurate high cadence optical variability analysis of AGN.

\begin{acknowledgments}
This work is supported by NASA grant number 80NSSC22K0741. KLS gratefully acknowledges the staff of the K2 Guest Observer office at NASA Ames, especially Christina Hedges, for their assistance and advice in adapting the matrix regression methods. LFS acknowledge the financial support of the Swiss National Science Foundation SNSF. LFS acknowledges Emily O. Garvin with her support for the choice and application of statistical tests. All of the data presented in this paper were obtained from the Mikulski Archive for Space Telescopes (MAST) at the Space Telescope Science Institute. The specific observations analyzed can be accessed via \dataset[DOI: 10.17909/k6mn-kp63]{https://doi.org/10.17909/k6mn-kp63}. Finally, we thank the anonymous referee for comments that substantially improved the clarity of the manuscript, as well as led to modifcations of \texttt{quaver} which will improve the user's ability to assess the validity of the fitting process.
\end{acknowledgments}

%

\vspace{5mm}
\facilities{TESS \citep{Ricker2015}, ZTF \citep{Masci2019}}


\software{Quaver (this work), TESSCut \citep{Brasseur2019}, Astropy (\citealt{Astropy2013,Astropy2018}), Astroquery \citep{Ginsburg2019}, Matplotlib (\citealt{Hunter2007}), Lightkurve (\citealt{Lightkurve2018}), EzTao (\citealt{Yu2022}).}
          
\vspace{2mm}         
{\large{\it{Authors contributions:}}} KLS developed and tested the Quaver tool, performed the flares analysis and wrote part of the manuscript. LAS extracted the TESS light curves of Pictor A, performed the variability analysis and wrote part of the manuscript. The authors collaborated throughout the project by discussing the possible analyses and the obtained results.



\appendix

\section{Handling of Scattered Light Background}
\label{app:scattered_light}

By far the dominant trend in uncorrected TESS light curves is the contribution from the scattered light from the moon and the Earth, which can leak into the FOV of the detector and varies significantly over the course of a given Sector. 

In the Simple Hybrid method, this light is removed using a simple subtraction of the light curve of the faint/empty pixels. 

\begin{figure*}
\centering
\includegraphics[width=\textwidth]{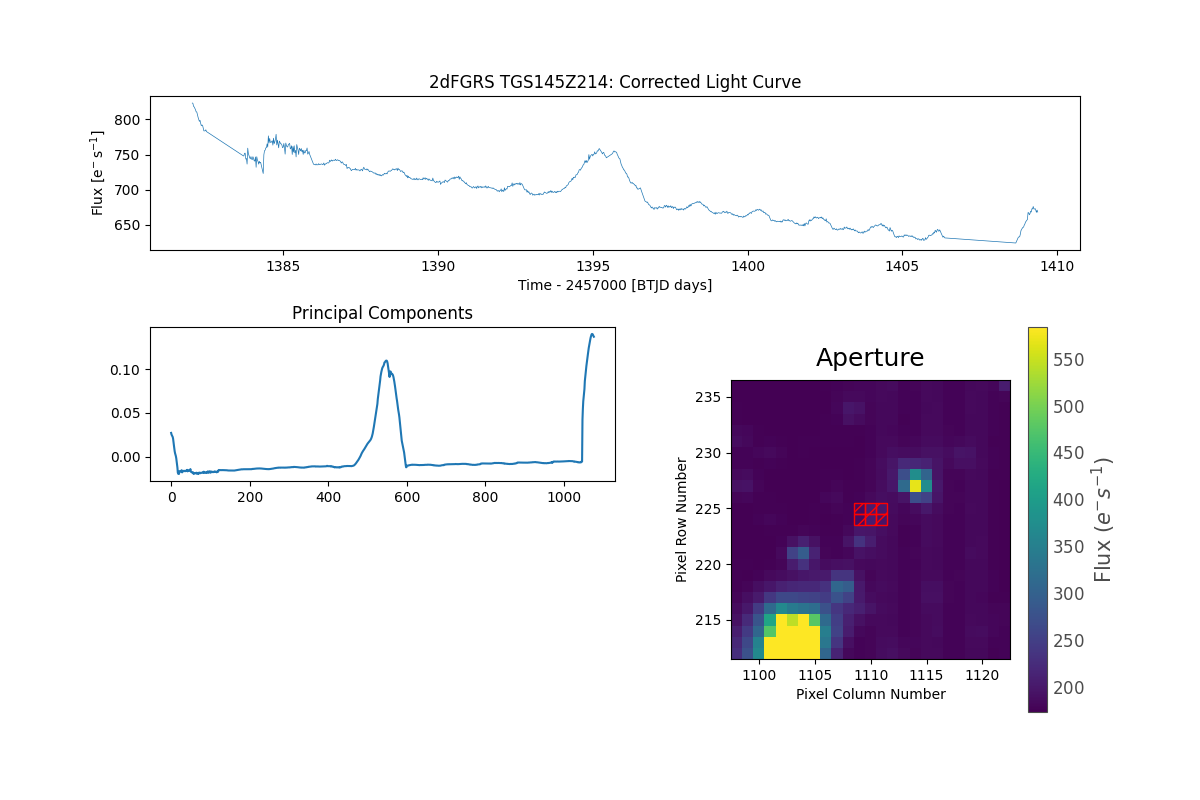}
\caption{TESS Sector~3 light curve of the non-AGN galaxy 2dFGRS~TGS145Z214, restricted to a one-component PCA analysis with insufficient parameters for background and contaminant correction.}
\label{fig:contam_galaxy_bad}
\end{figure*}

\begin{figure*}
\centering
\includegraphics[width=\textwidth]{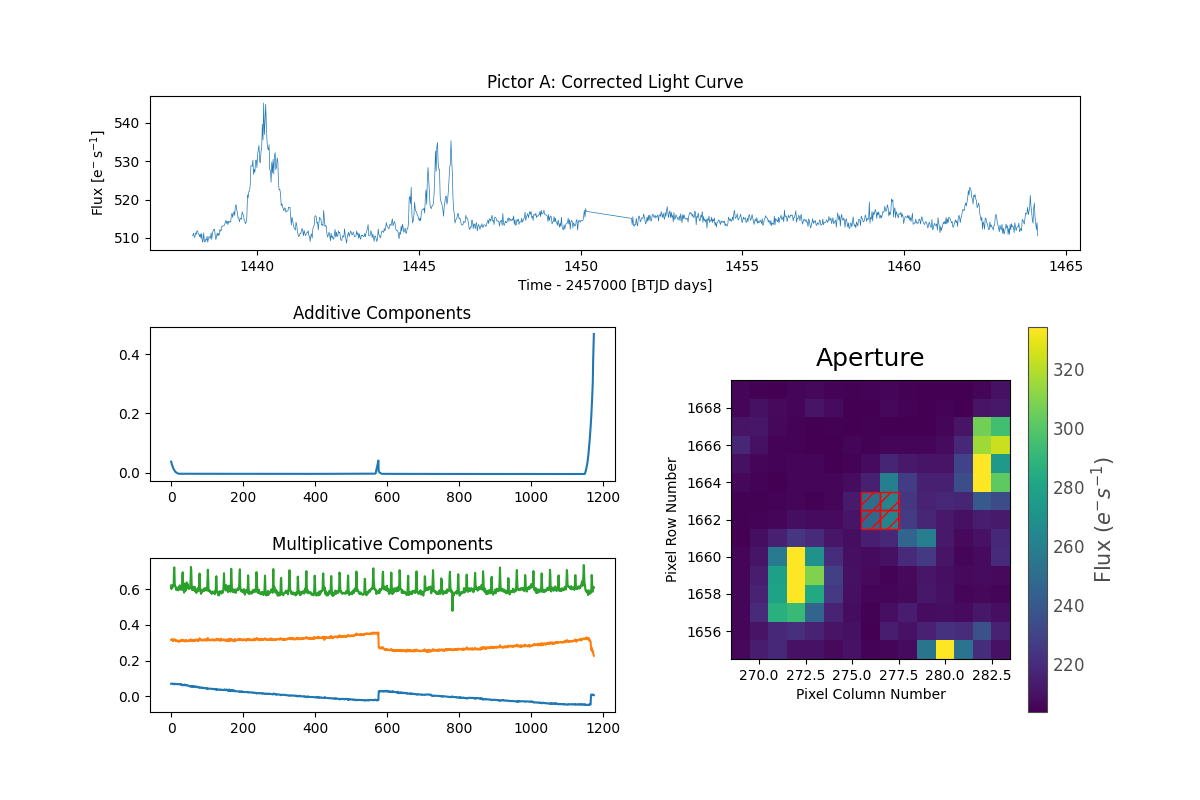}
\caption{TESS Sector~5 light curve of Pictor~A, using the Full Hybrid method and restricted to only one additive PCA component.}
\label{fig:scattered_light_example}
\end{figure*}

\begin{figure}
     \subfloat{
         \includegraphics[width=0.4\textwidth]{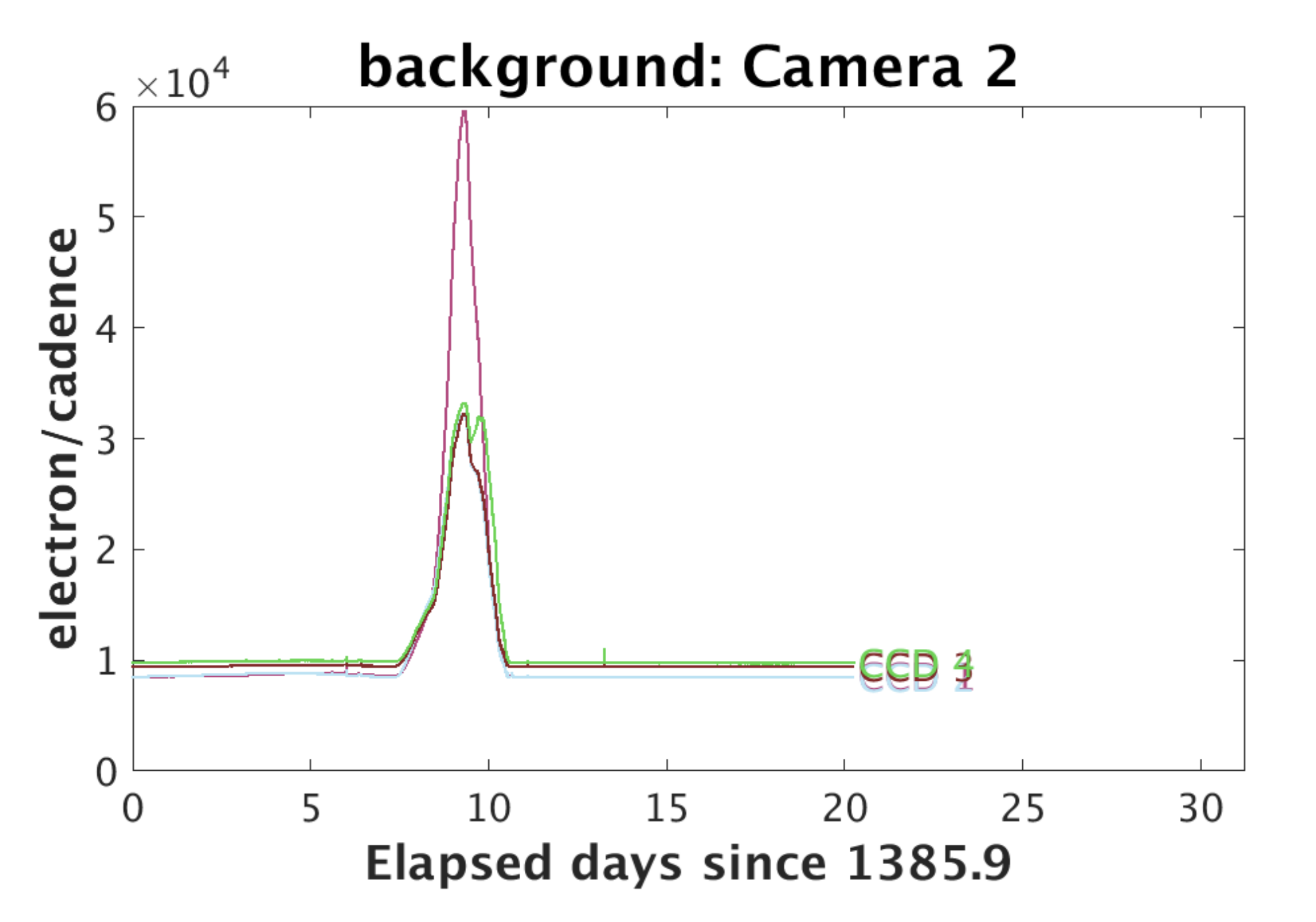}}

     \subfloat{
         \includegraphics[width=0.4\textwidth]{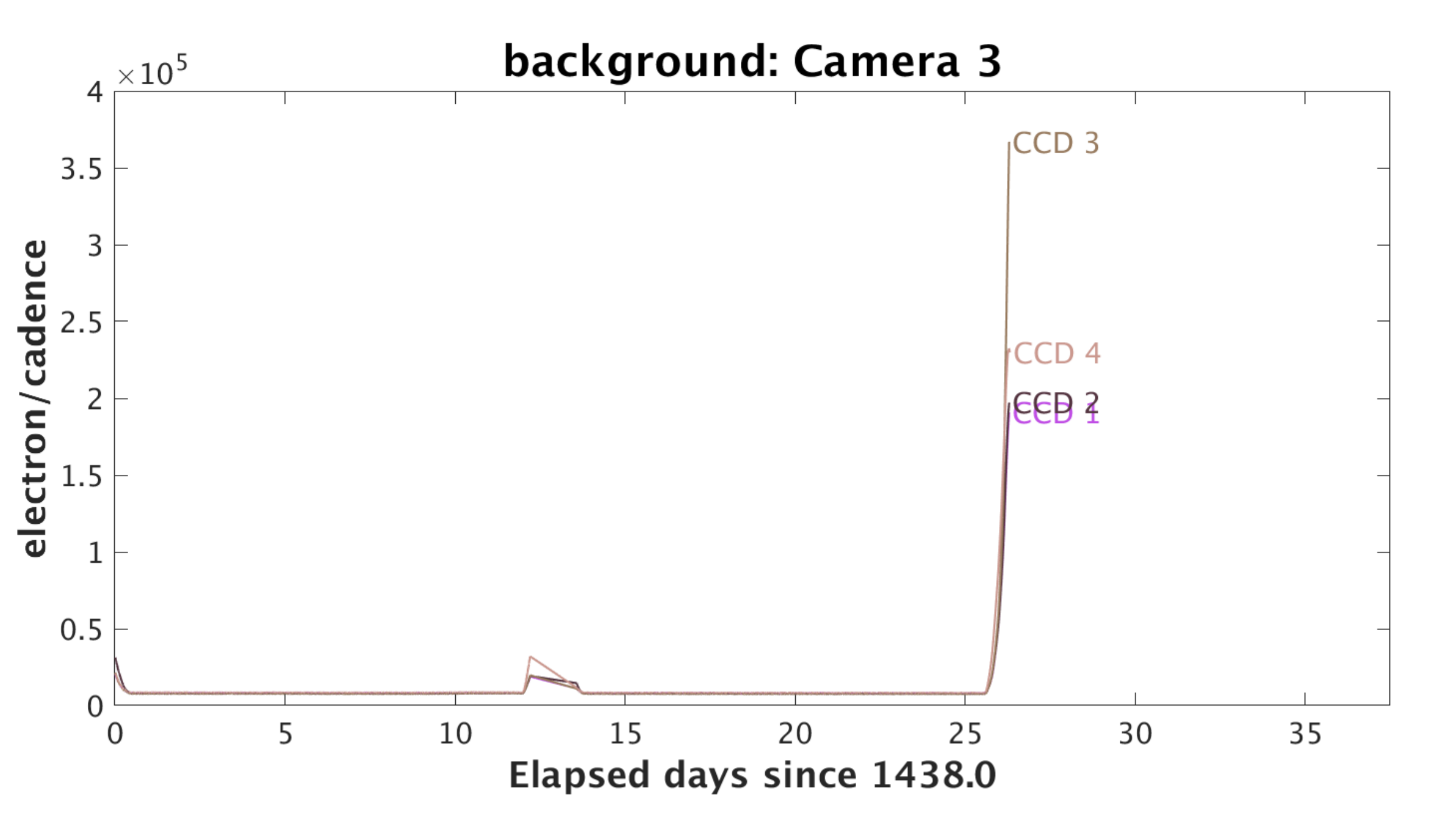}}
        \caption{Plots of the scattered background light over time for Sector 3, Camera 2 (left) and Sector 5, Camera 3 (right), from the TESS Data Release Notes (\url{https://archive.stsci.edu/tess/tess_drn.html}). Differences in plot formatting are due to differences as presented in the Sector 3 and 5 release note documents.}
        \label{fig:tess_drn_bgs}
\end{figure}

In the PCA and full-hybrid methods, the scattered light curve is always the dominant (first) principal component, and is removed in the matrix regression as that component. We demonstrate this in Figures~\ref{fig:contam_galaxy_bad} and \ref{fig:scattered_light_example}, where we show two example sectors: the extracted light curve of the inactive galaxy 2dFGRS~TGS145Z214 (see Section~\ref{app:contamination}) using the PCA method with only one PCA chosen, and the the extracted light curve of Pictor~A using the Full Hybrid method with only one additive PCA allowed. If this is compared to Figure~\ref{fig:tess_drn_bgs}, it is clear that the primary contribution to both light curves is the same as the scattered light trend for the appropriate camera and sector.

\newpage
\section{Contamination from Variable Sources}
\label{app:contamination}
The large pixel response function (PRF) of TESS means that even at apparently large distances from the target source, variable stars of sufficient magnitude can contribute significantly to the flux in the target aperture. In some cases, it is sufficient to cull one's sample of all targets in close proximity to potential contaminants, as in the recent work selecting AGN using TESS variability by \citet{Treiber2022}. However, the number of AGN below TESS's magnitude limit is already only a few thousand, and those with long light curves number in the hundreds. It is therefore desirable to be able to correct for nearby source contamination.

\begin{figure}
\centering
\includegraphics[width=0.45\textwidth]{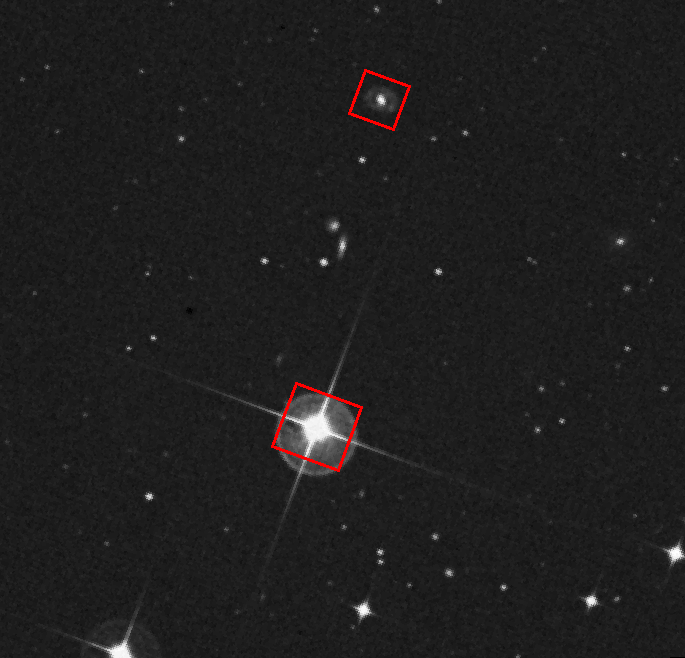}
\caption{Map of the variable star AP~Sculptor and the nearby galaxy 2dFGRS~TGS145Z214, each shown in a red box. The sources are separated by 4.79~arcmin.}
\label{fig:contam_starmap}
\end{figure}

\begin{figure*}
\centering
\includegraphics[width=\textwidth]{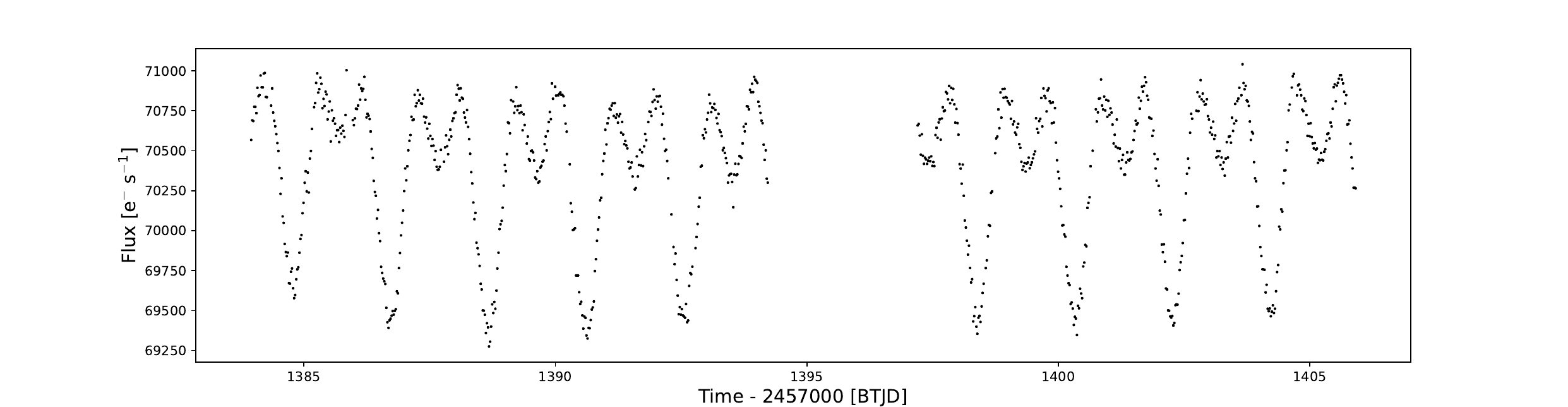}
\caption{TESS Sector~3 light curve of the variable star AP~Sculptor, extracted with \texttt{Quaver}.}
\label{fig:contam_stellar_lc}
\end{figure*}

\begin{figure*}
\centering
\includegraphics[width=\textwidth]{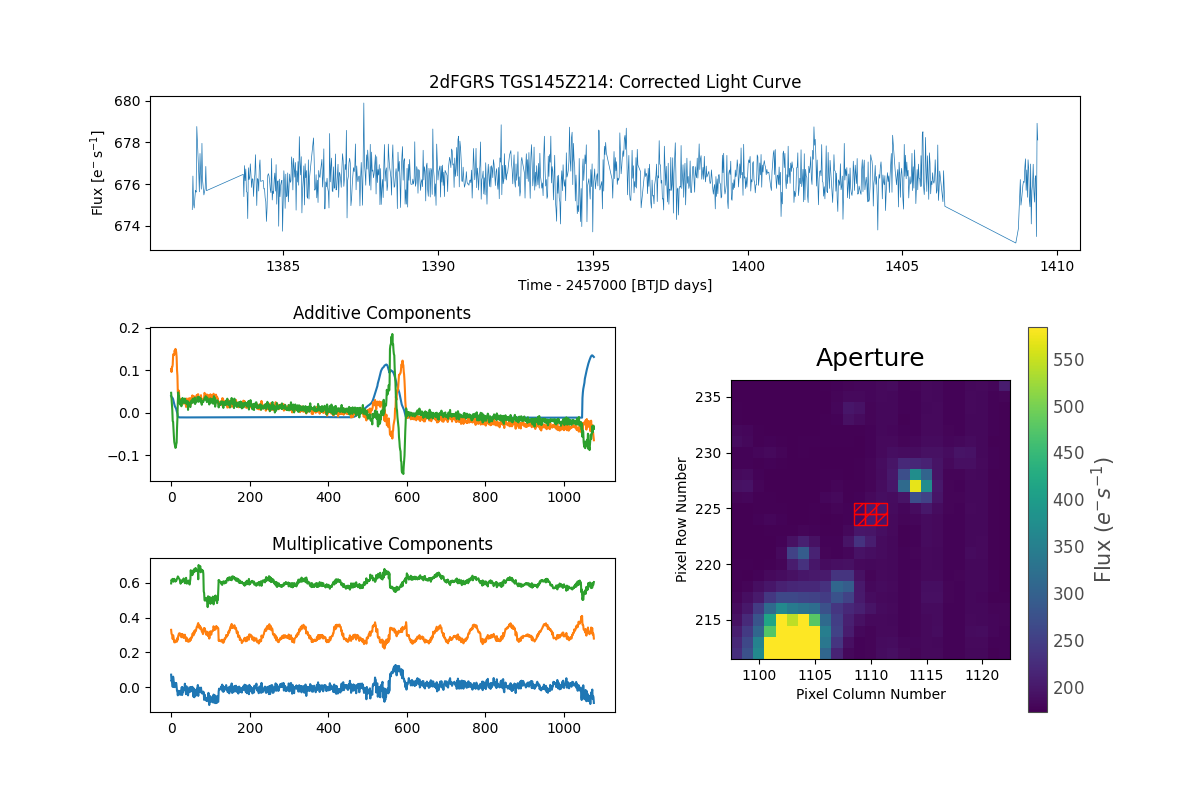}
\caption{TESS Sector~3 light curve of the non-AGN galaxy 2dFGRS~TGS145Z214, extracted with \texttt{Quaver}'s full hybrid method.}
\label{fig:contam_galaxy_flat}
\end{figure*}
The custom aperture selection is the first line of defense, allowing the user to manually exclude source from nearby bright sources and use the DSS overlay to ascertain the proximity of even faint nearby stars. However, it is not obvious by eye which stars may still be contributing to flux inside the target aperture. 

The Hybrid methods mentioned above provide an important safeguard against strongly variable signals: all pixels containing a source (i.e., containing a source above the user-defined threshold) are background-corrected, using either subtraction in the case of the Simple Hybrid method or using a separate regression matrix in the Full Hybrid method. Then, these background-corrected sources are \emph{themselves} fitted by principal components, to create the second correction matrix. Any strongly variable sources will contribute to these components, and this contaminant variability will be corrected for. 

To demonstrate the effectiveness with which \texttt{Quaver} removes these trends, we extract the light curve of the bright ($V=8.4$) $\alpha^2$~CVn variable star AP~Sculptor \citep{Balona2022}, and the light curve of the nearby non-AGN (and thus non-variable) galaxy 2dFGRS~TGS145Z214 ($V=15.98$). The objects are separated by 4.8~arcmin, and the star is very bright, with a flux distribution across many pixels. 
 
In Figure~\ref{fig:contam_starmap}, we show a map of the variable star and galaxy. In Figure~\ref{fig:contam_stellar_lc}, we show the result of a \texttt{Quaver} extraction of AP~Sculptor, to demonstrate its strong variability. In Figure~\ref{fig:contam_galaxy_flat} we show the \texttt{Quaver}-corrected light curve of 2dFGRS~TGS145Z214 (using the Full Hybrid method, with three principle components for the additive and multiplicative components).

The second component in the Multiplicative Components panel is contaminant variability AP~Sculptor. This has been removed from the galaxy light curve, which is flat, as expected. 
\newpage

\section{Light Curve Correction Figures}
\label{app:corrections}

Here we provide the output from Quaver which shows the light curves before and after correction, along with the final models derived by the matrix regression. This Appendix includes these plots for Star~1, Sector~4 and Star~3, Sector~6 from Figure~\ref{fig:ref_stars_lcs}, as well as for Pictor~A (except for Pictor~A Sector~5, which was shown in the main document). In each figure, the top pane shows the results of the Simple PCA fitting; the middle pane shows the results of The Simple Hybrid method by showing the uncorrected light curve and the subtracted background in the top panel, and the background-subtracted and corrected light curve with the removed multiplicative model in the bottom panel; and the bottom pane shows the results of the Full Hybrid method.

\begin{figure}
\centering
\begin{minipage}{.45\textwidth}
  \centering
  \includegraphics[width=\linewidth]{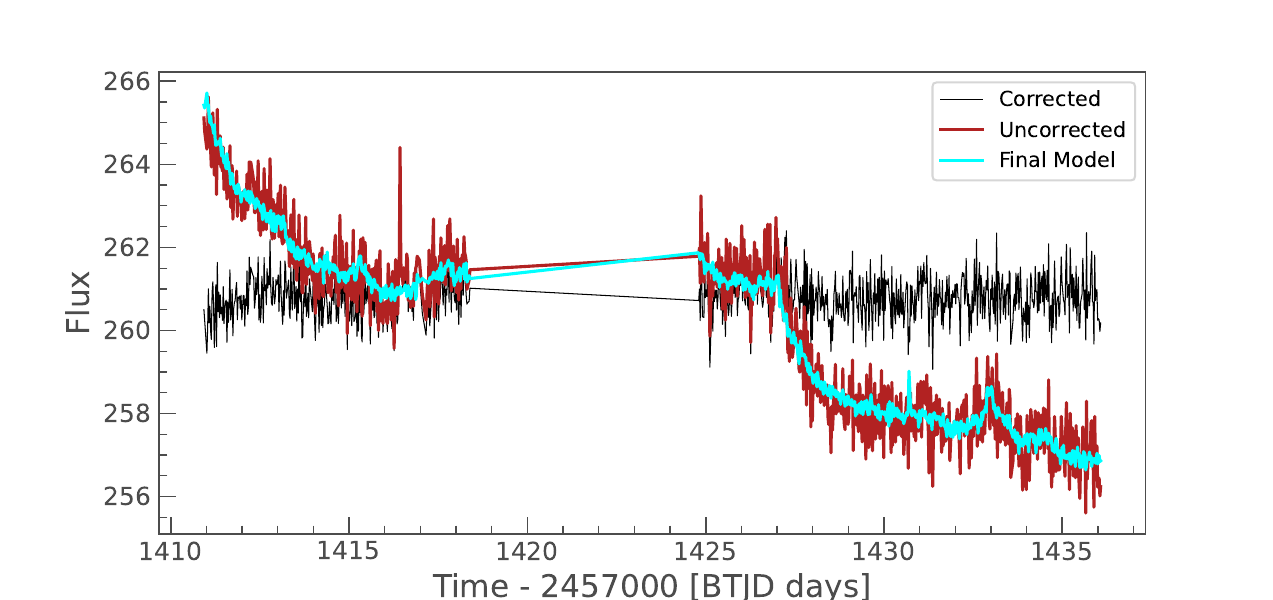}
    \includegraphics[width=\linewidth]{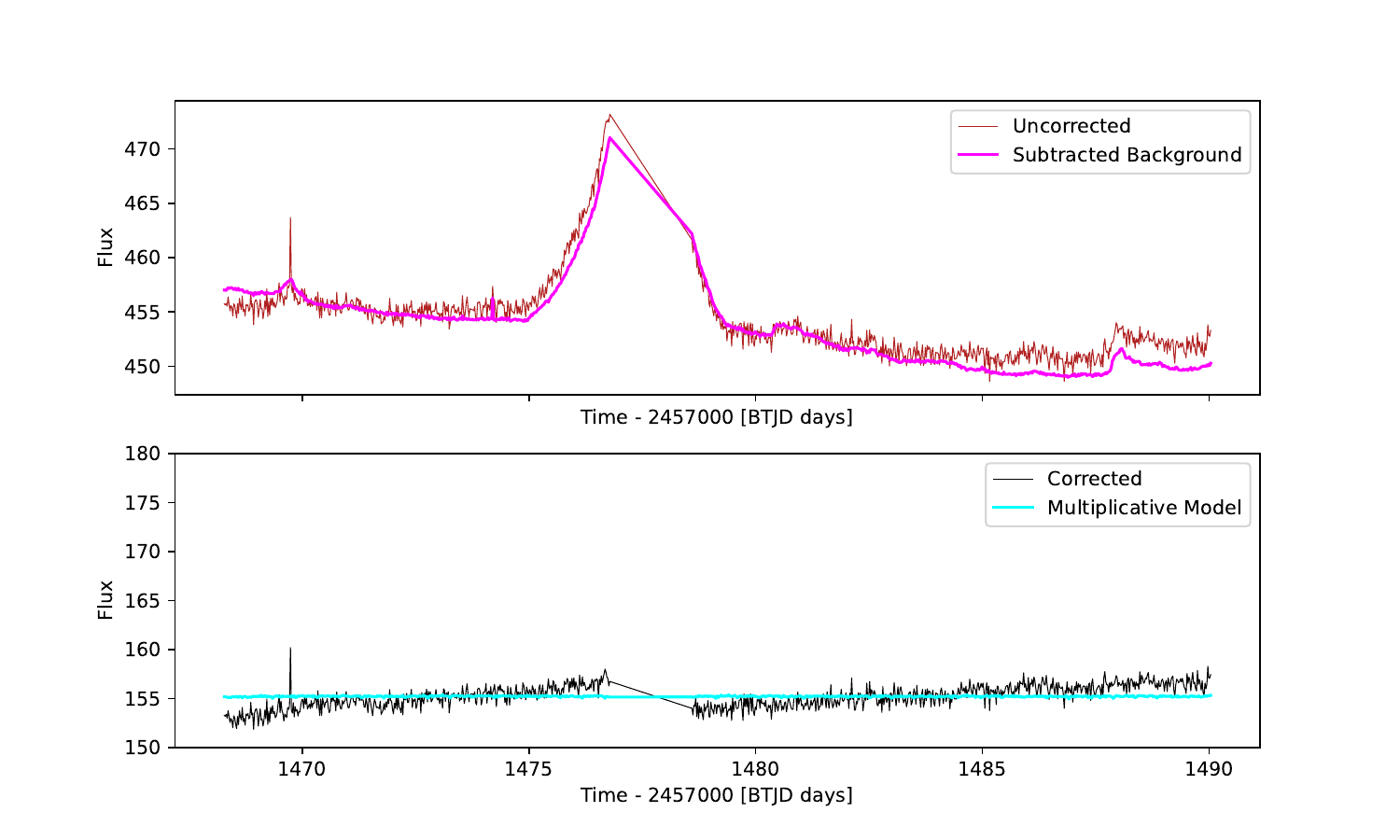}
    \includegraphics[width=\linewidth]{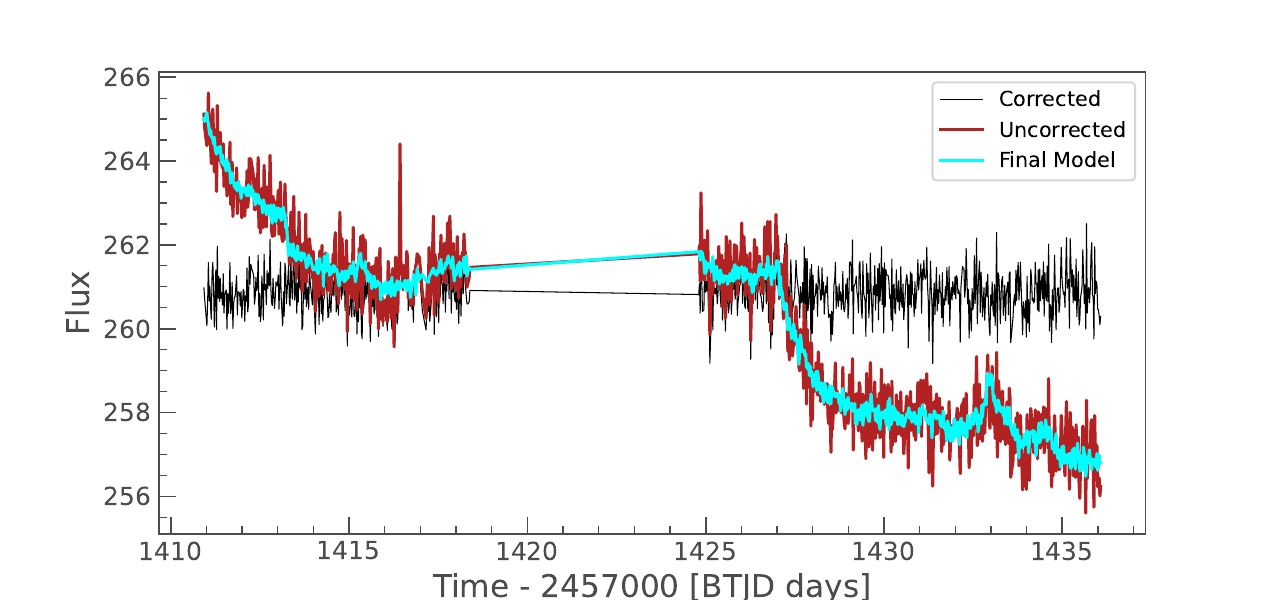}
  \caption{Star 1, Sector 4.}
  \label{fig:test1}
\end{minipage}%
\begin{minipage}{.45\textwidth}
  \centering
  \includegraphics[width=\linewidth]{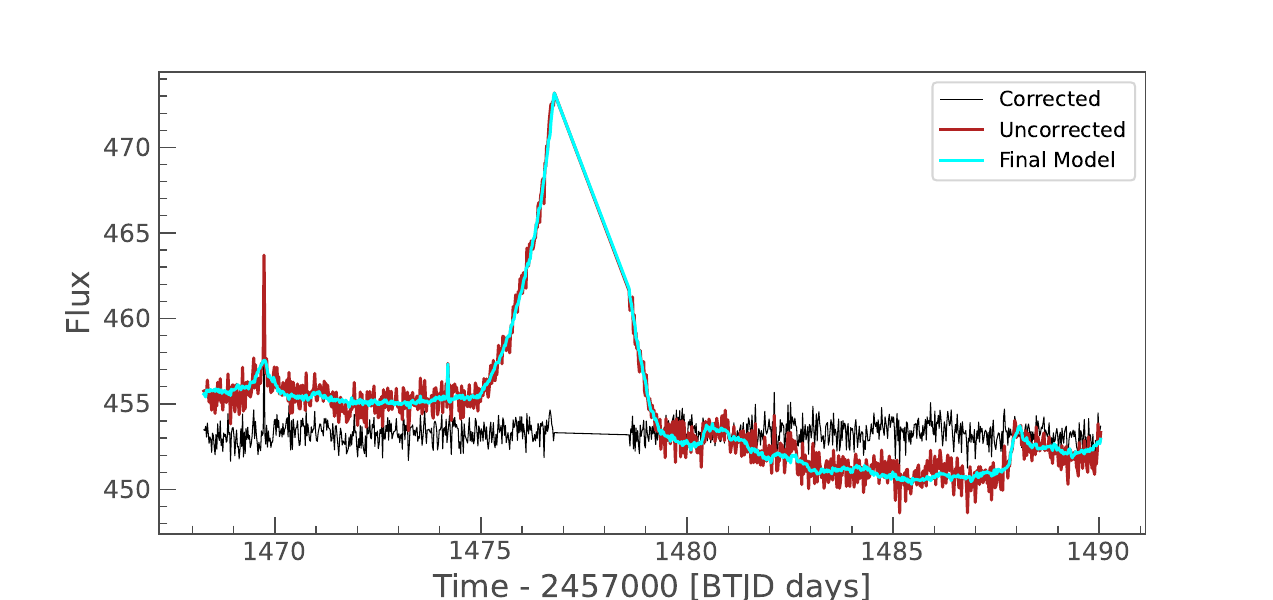}
    \includegraphics[width=\linewidth]{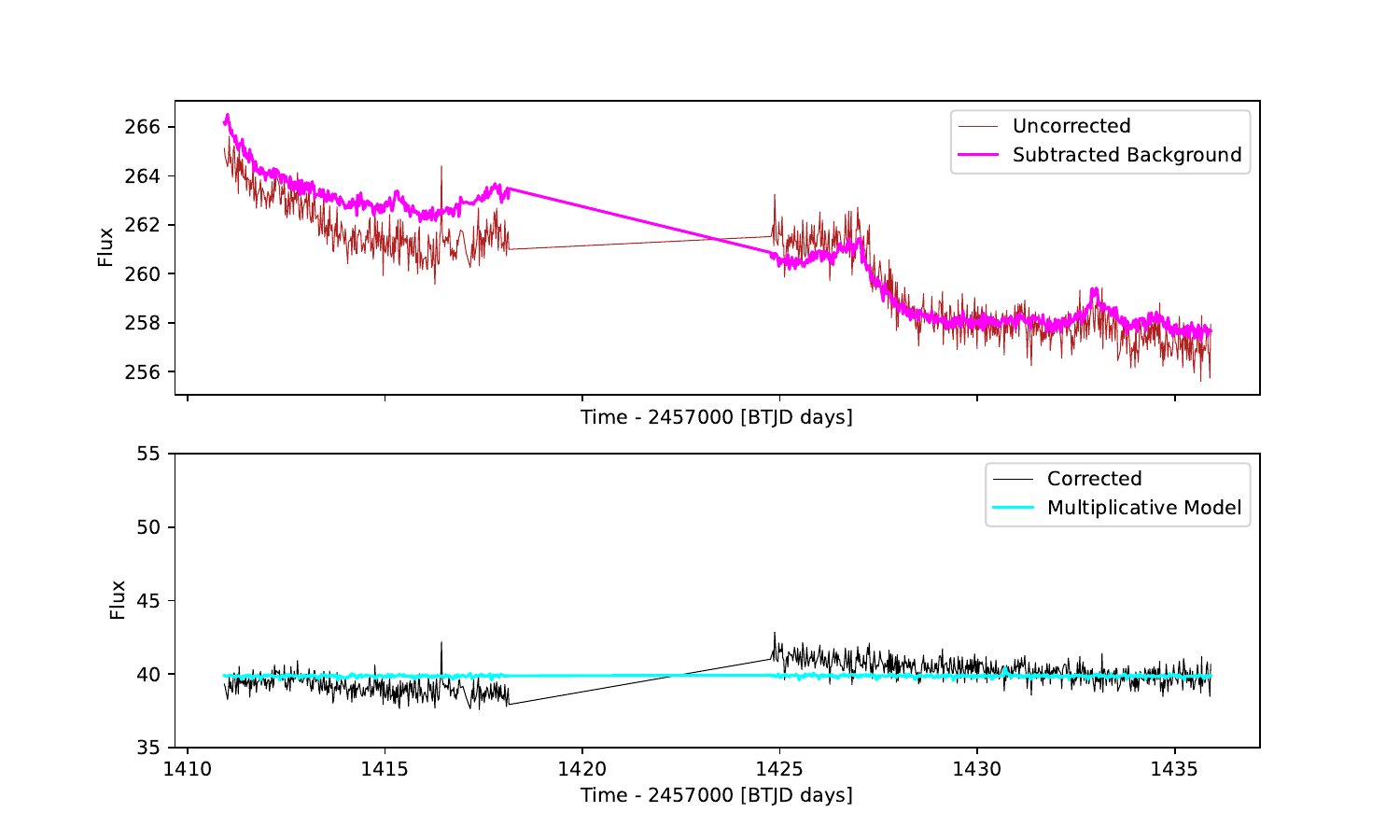}
 \includegraphics[width=\linewidth]{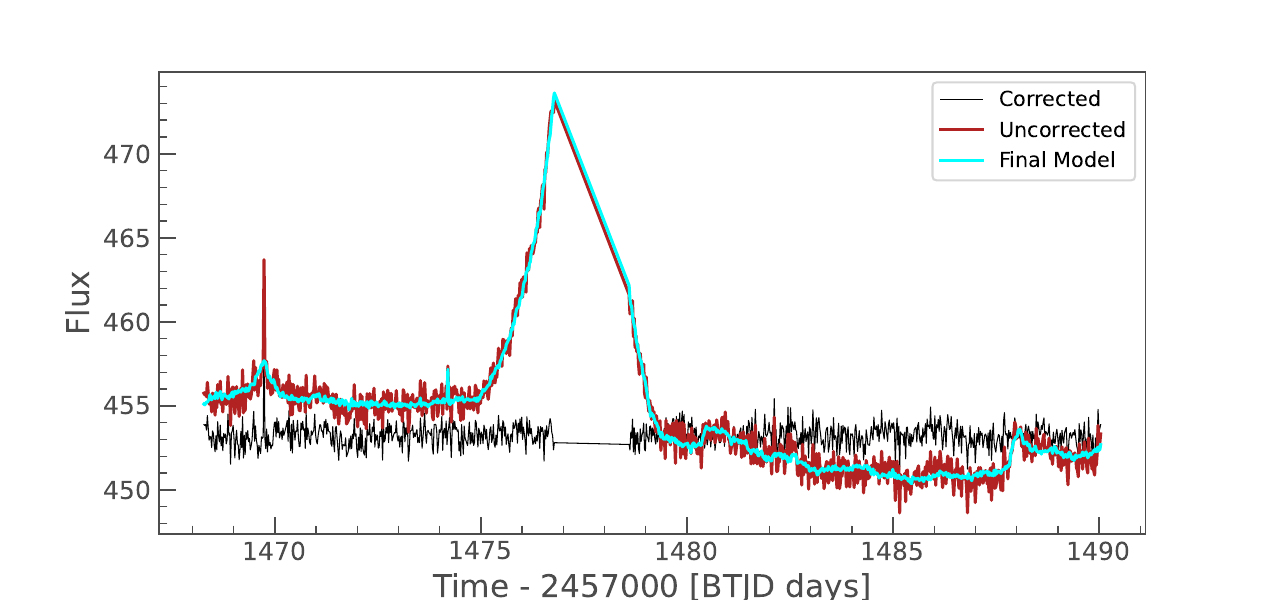}
  \caption{Star 3, Sector 6.}
  \label{fig:test_stars_fitdiag}
\end{minipage}
\end{figure}

\begin{figure}
\centering
\begin{minipage}{.45\textwidth}
  \centering
  \includegraphics[width=\linewidth]{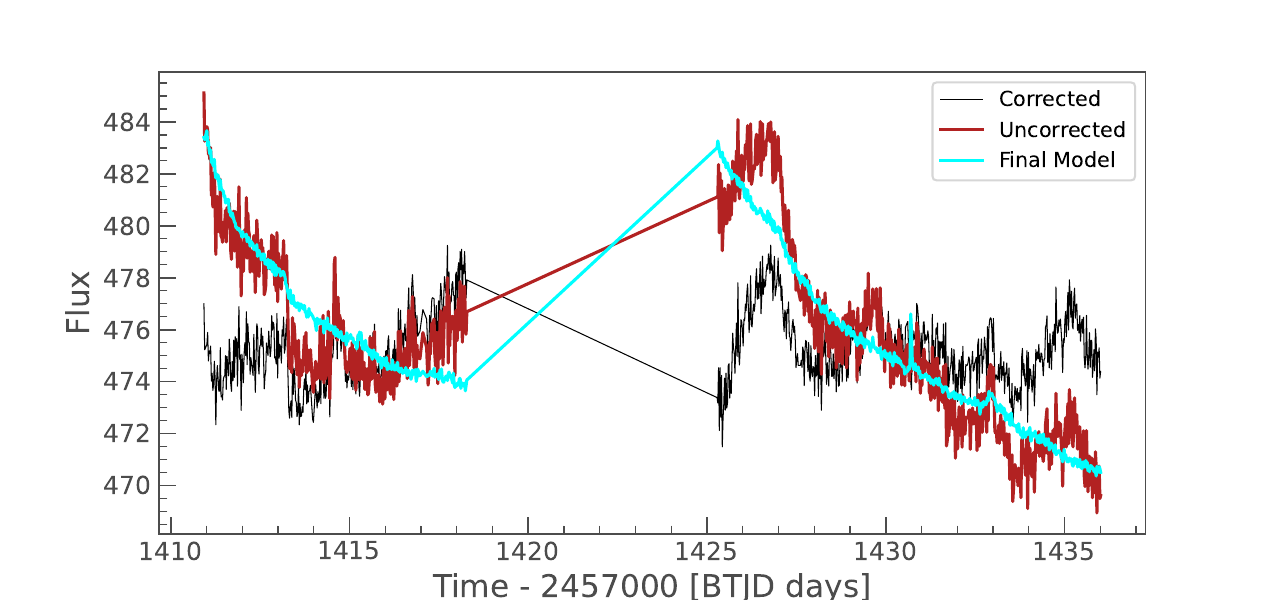}
    \includegraphics[width=\linewidth]{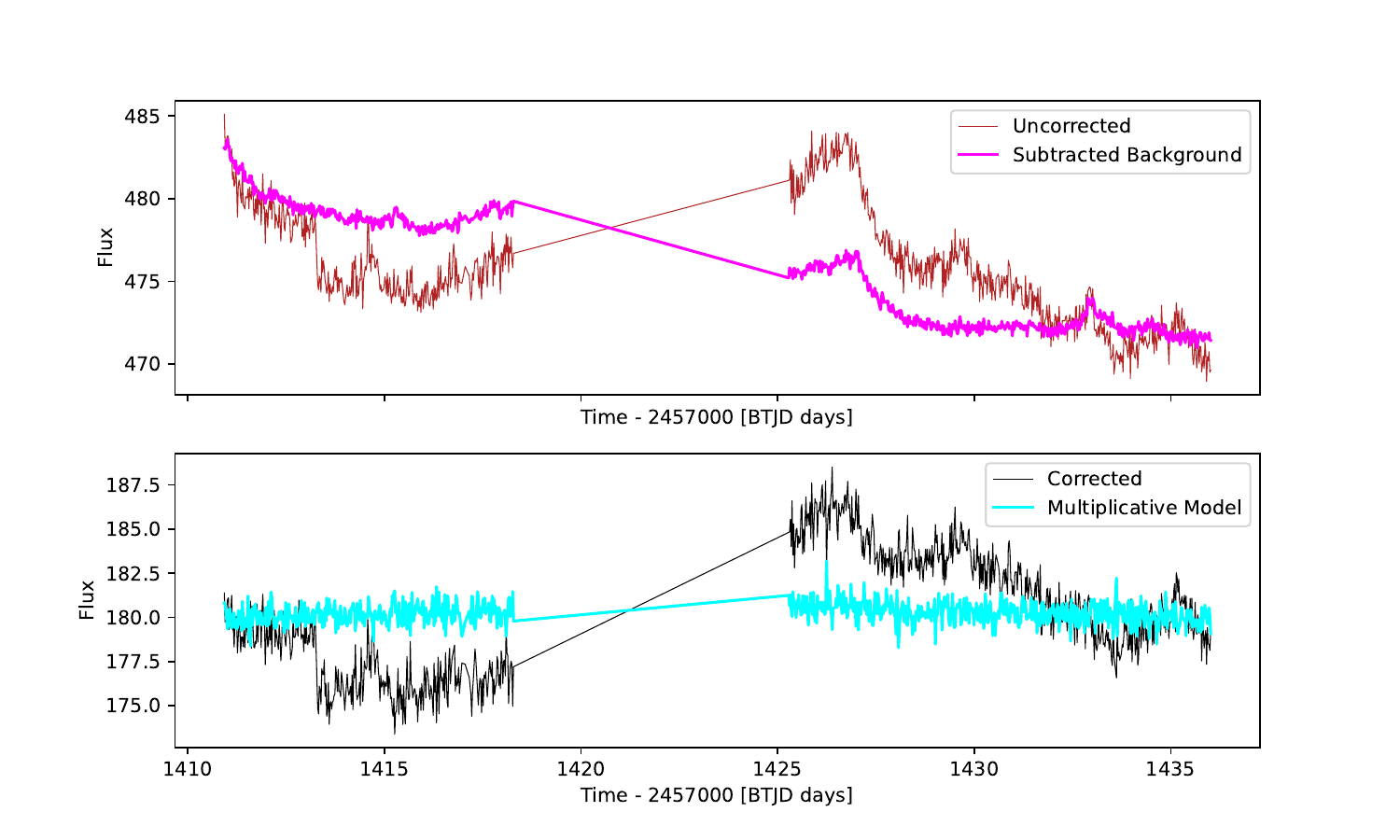}
    \includegraphics[width=\linewidth]{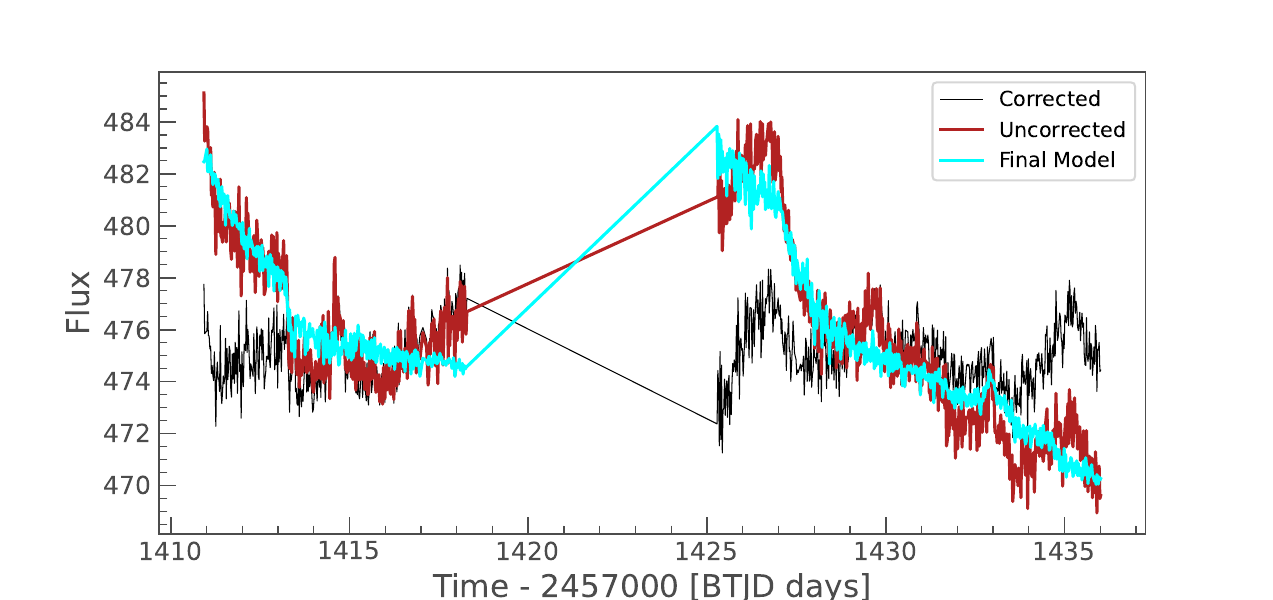}
  \caption{Pictor A, Sector 4.}
  \label{fig:test1}
\end{minipage}%
\begin{minipage}{.45\textwidth}
  \centering
  \includegraphics[width=\linewidth]{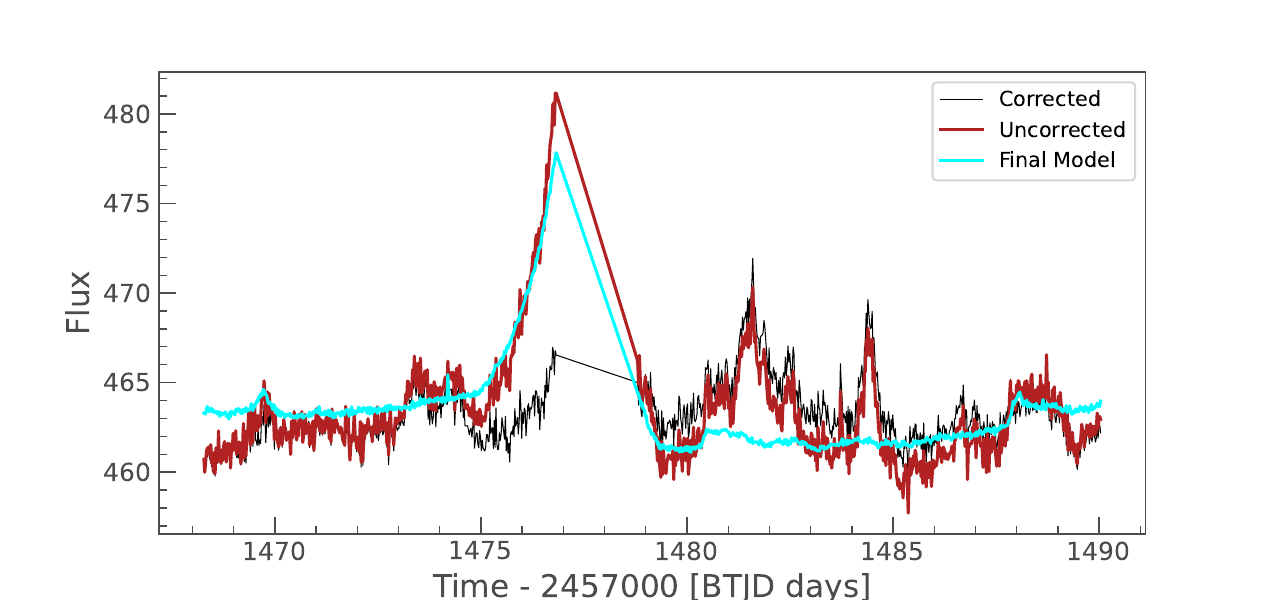}
    \includegraphics[width=\linewidth]{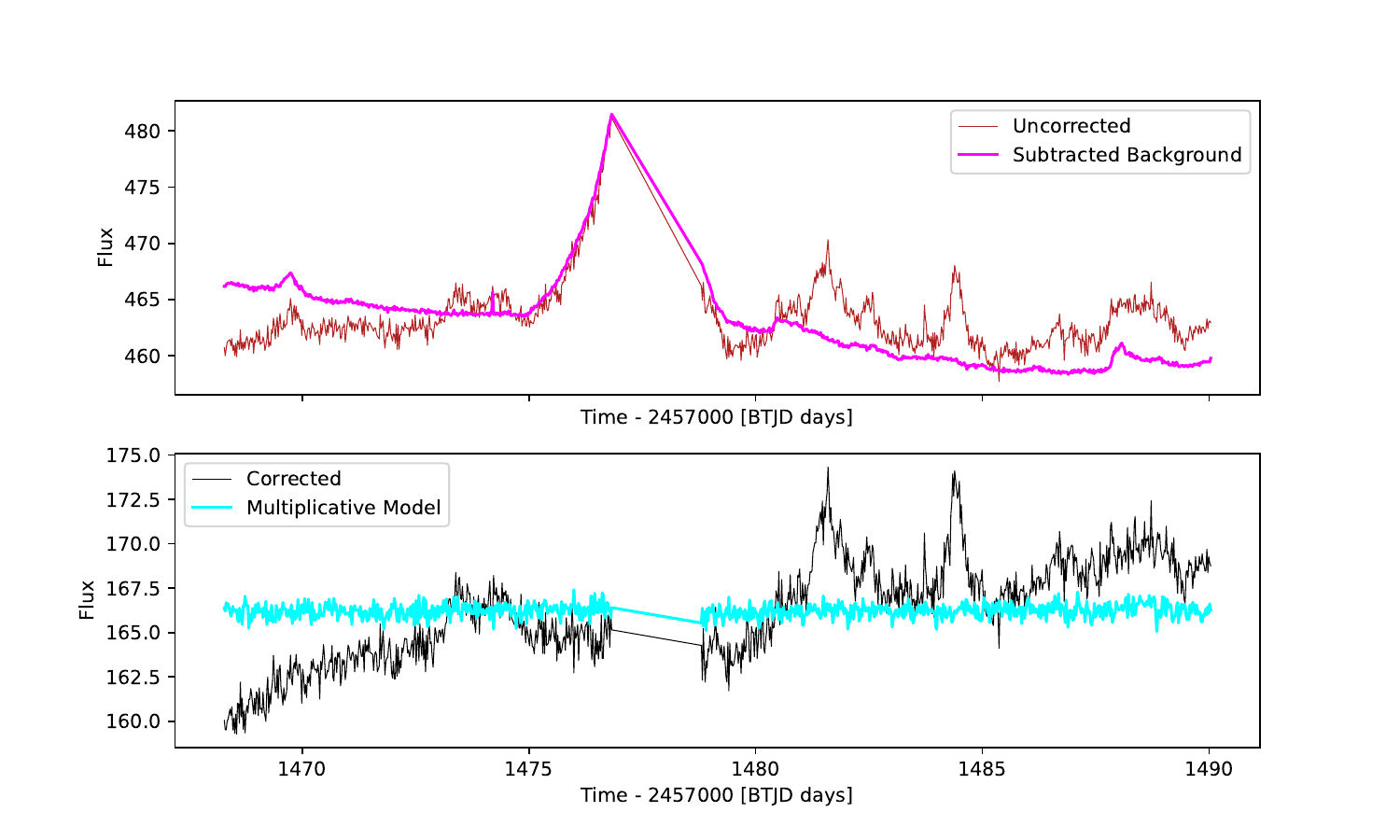}
 \includegraphics[width=\linewidth]{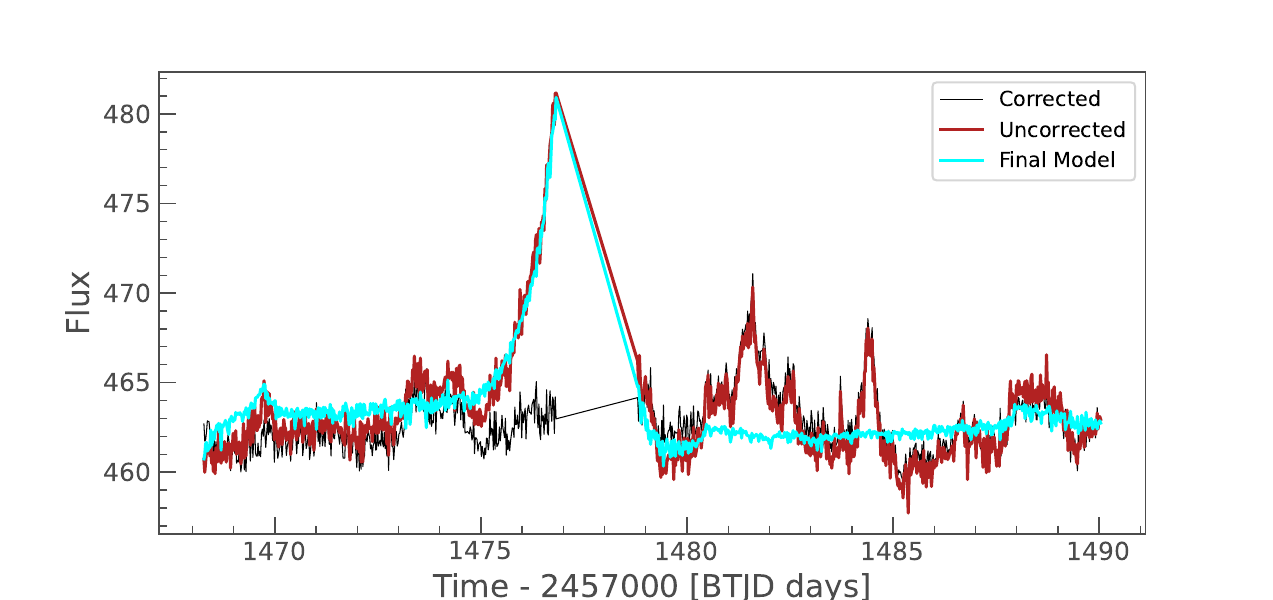}
  \caption{Pictor A, Sector 6.}
  \label{fig:test2}
\end{minipage}
\end{figure}

\begin{figure}
\centering
\begin{minipage}{.45\textwidth}
  \centering
  \includegraphics[width=\linewidth]{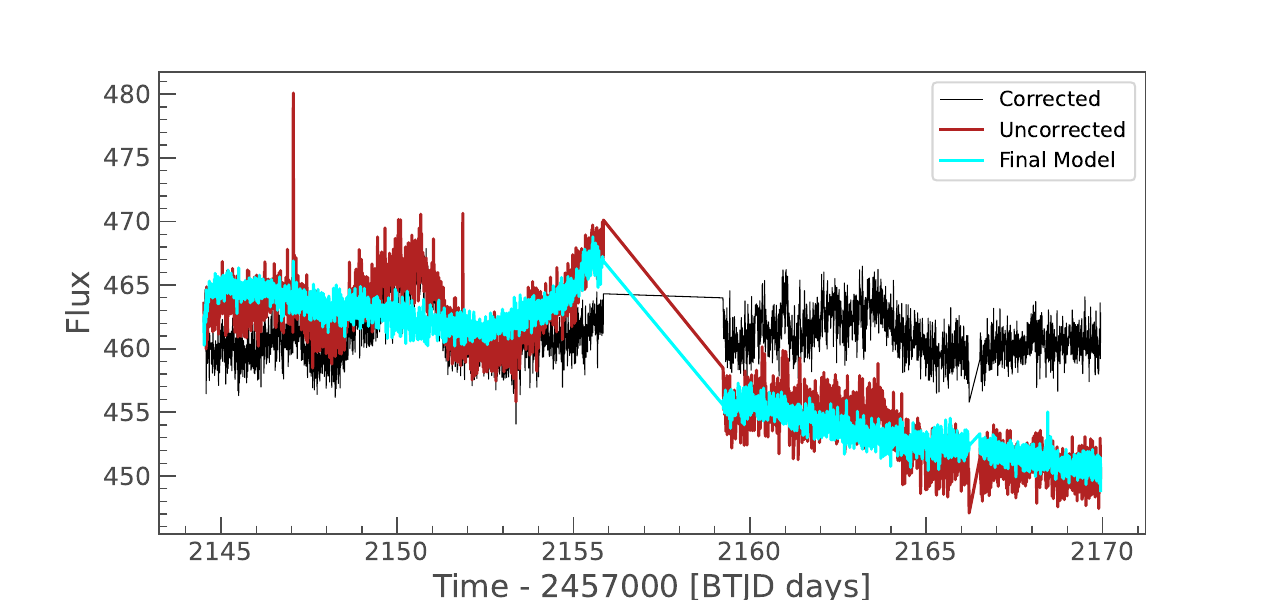}
    \includegraphics[width=\linewidth]{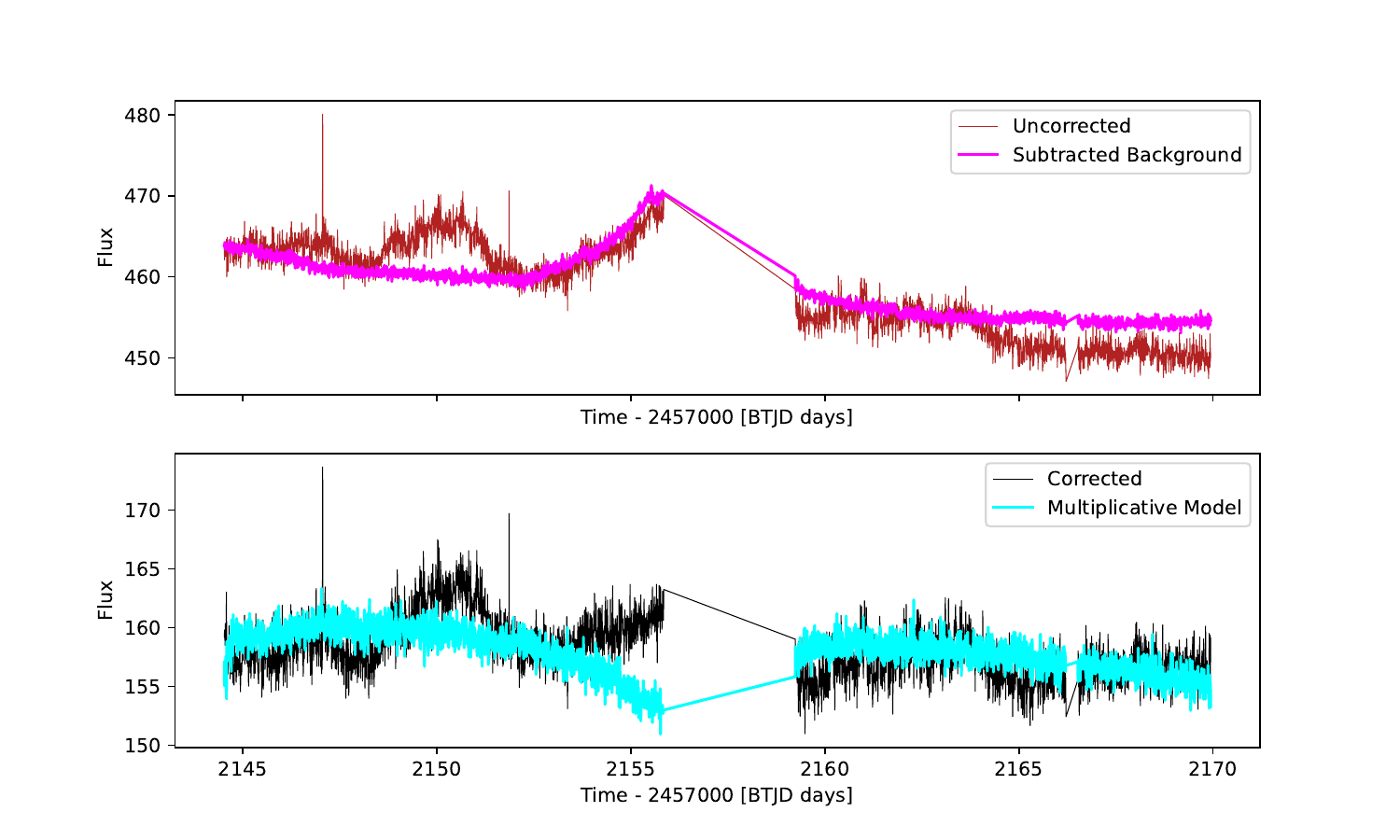}
    \includegraphics[width=\linewidth]{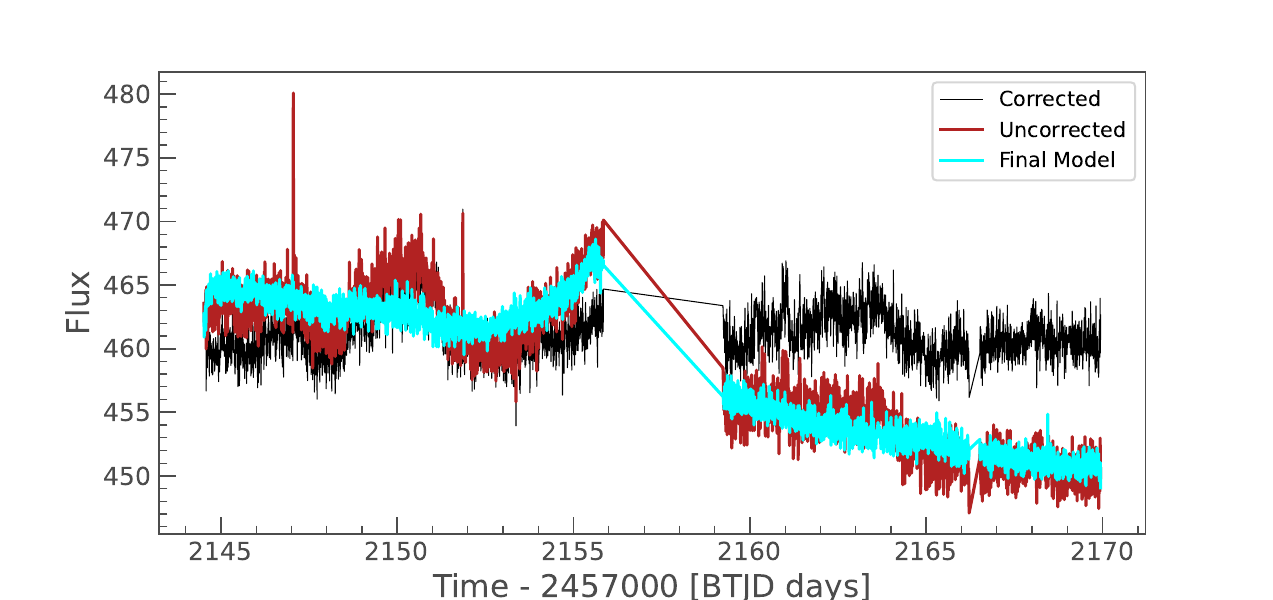}
  \caption{Pictor A, Sector 31.}
  \label{fig:test1}
\end{minipage}%
\begin{minipage}{.45\textwidth}
  \centering
  \includegraphics[width=\linewidth]{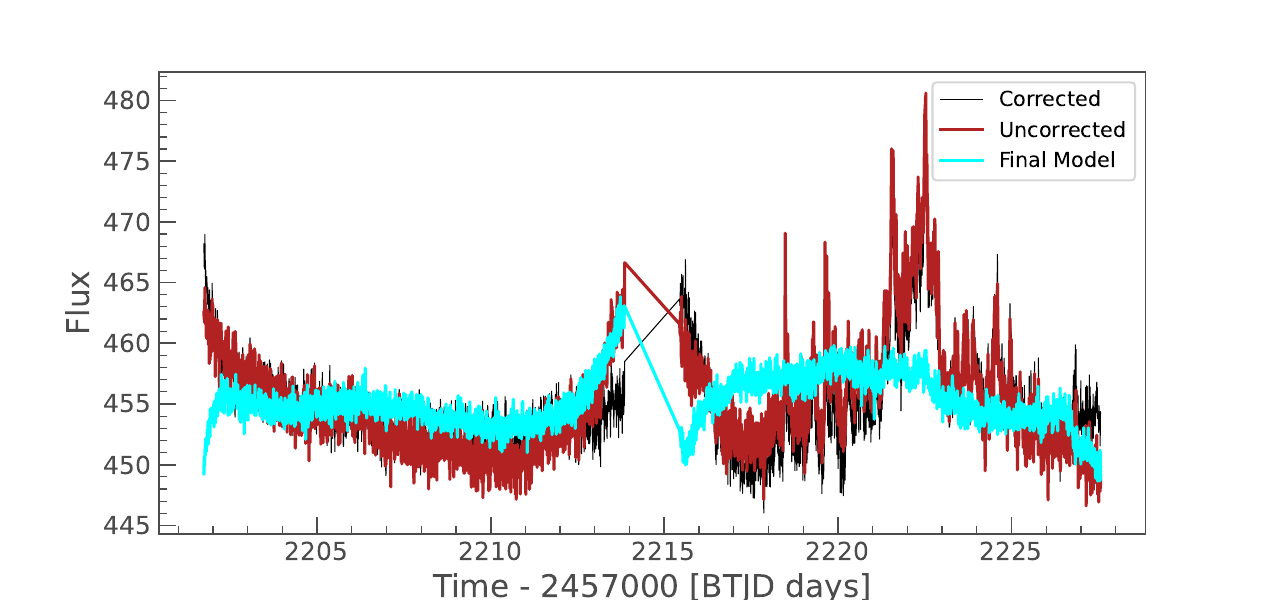}
    \includegraphics[width=\linewidth]{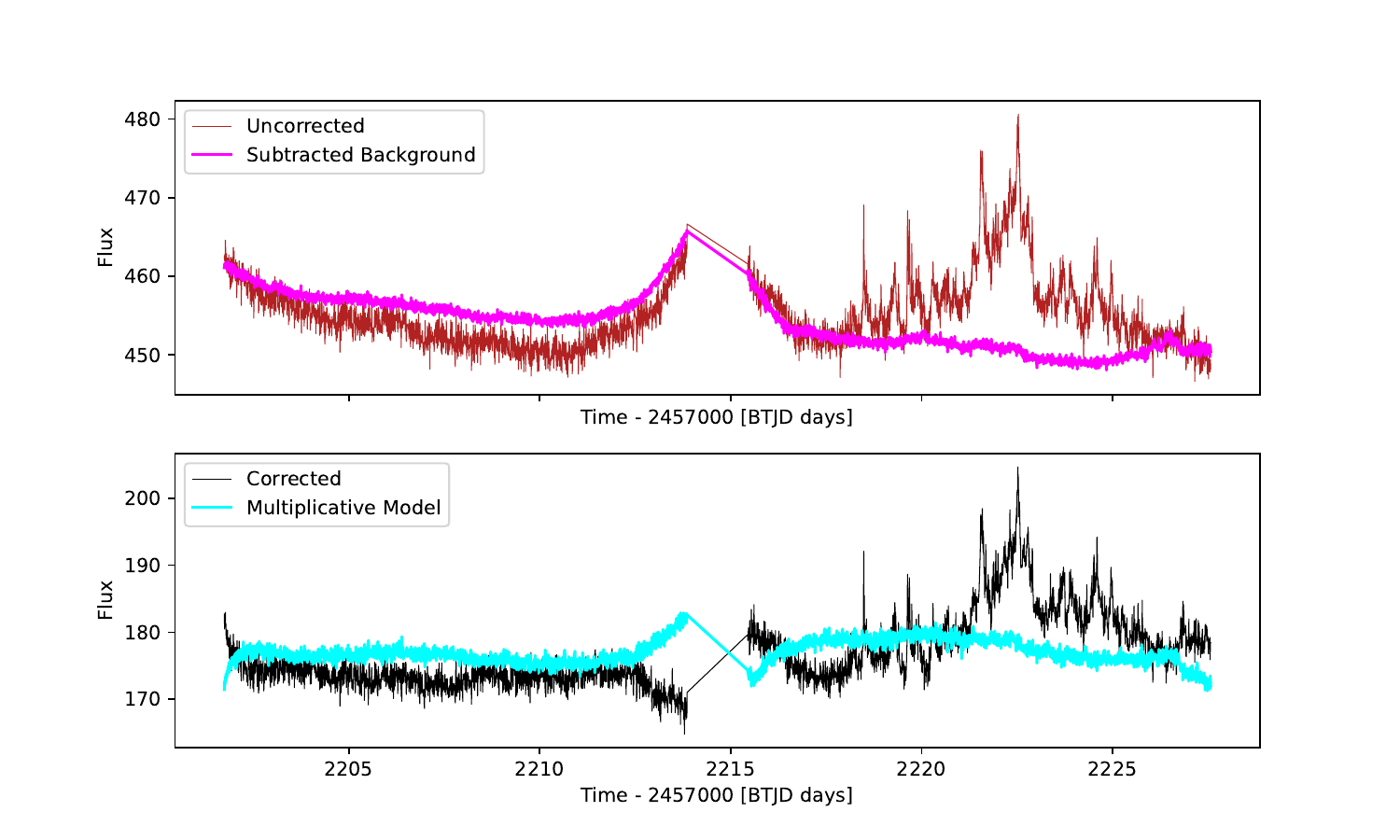}
 \includegraphics[width=\linewidth]{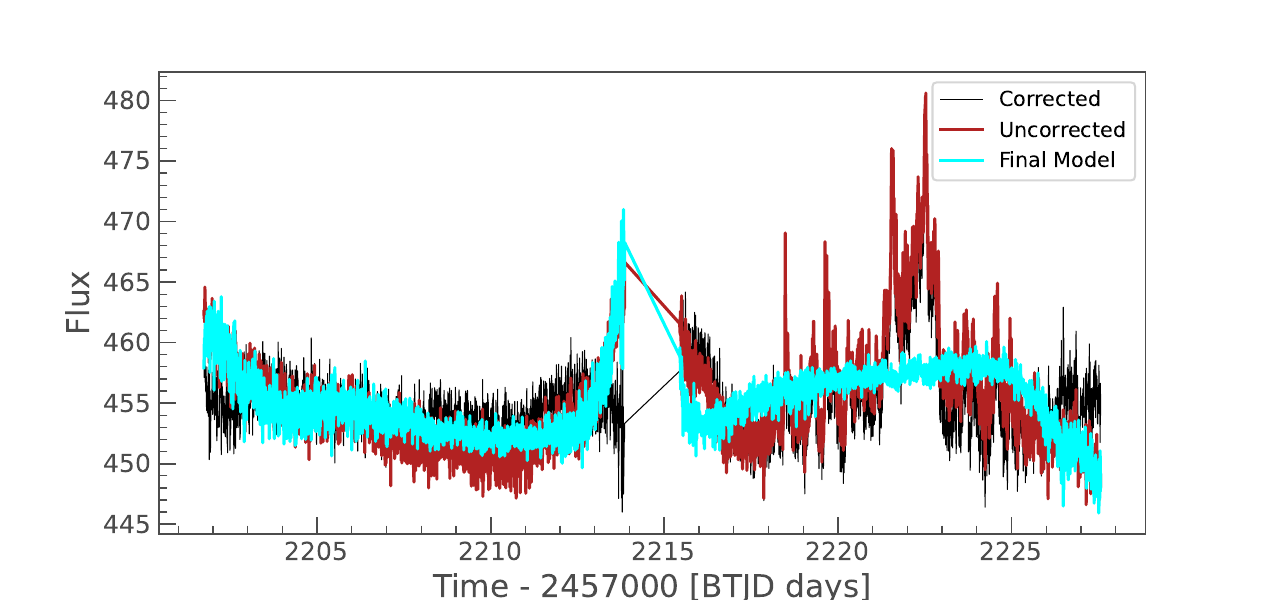}
  \caption{Pictor A, Sector 33.}
  \label{fig:test2}
\end{minipage}
\end{figure}

\newpage
\section{Damped Random Walk fit}\label{app:DRW}
\begin{figure*}
\centering
\includegraphics[width=\textwidth]{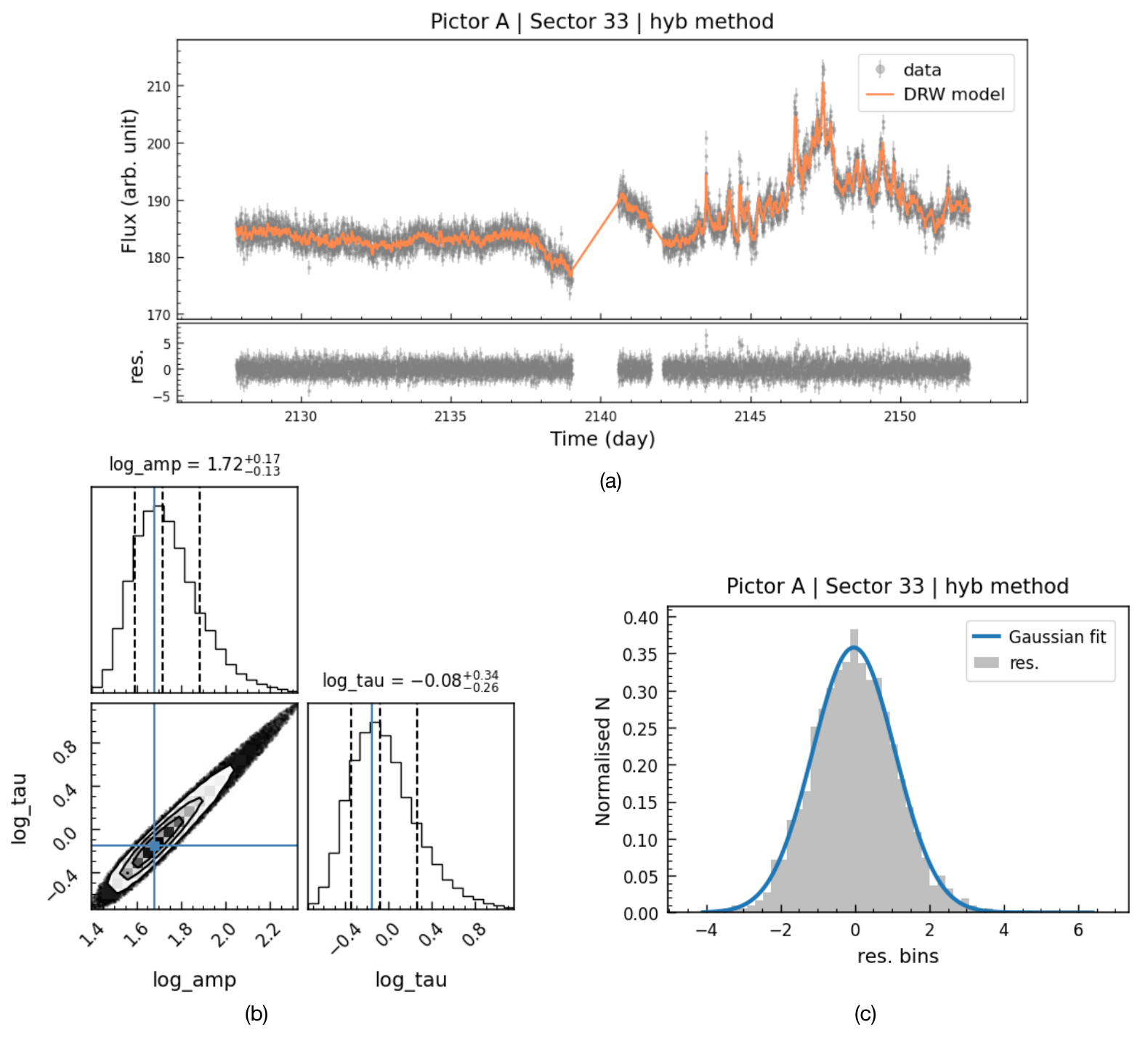}
\caption{Example of DRW fit performed for the light curve of Sector 33. \textit{Panel (a), top}: rest frame TESS light curve for Sector 33 (gray) and best-fit DRW model (orange). \textit{Panel (a), bottom}: residuals from the DRW fit. As we can see in \textit{panel (c)}, the histogram of the residuals (gray) follows a Gaussian distribution (blue). \textit{Panel (b)}: Posteriors distribution from the DRW fit preformed with \texttt{EzTao}. Both parameters, i.e. amplitude and $\tau{_{\rm DRW}}$, are well constrained. See Section \ref{sec:per} for details.}
\label{fig:ex_fit_res}
\end{figure*}

Here we provide a few additional figures to support the tests carried out in Section~\ref{sec:per}. In Figure~\ref{fig:ex_fit_res}, we provide an example of a DRW fit result, with posterior distributions constraining the parameter, and a histogram showing that the residulals of the fit are normally distributed. This, along with the results of the Durbin-Watson test as described in Section~\ref{sec:per}, supports the statement that the DRW is an acceptable model for the observed variability behavior. 

\begin{figure*}
\centering
\includegraphics[width=0.8\textwidth]{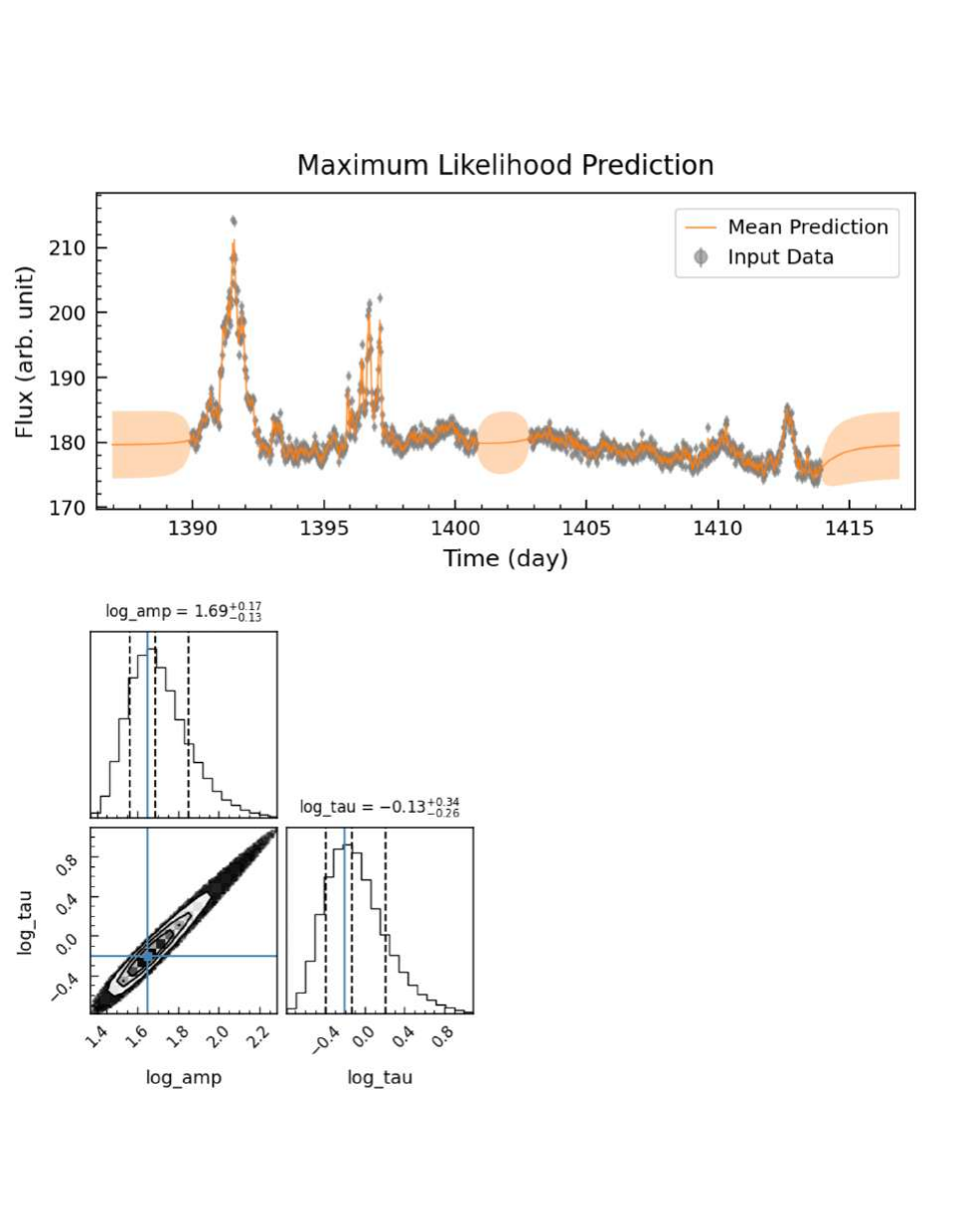}
\caption{Results of the DRW fit to Sector~5 data as observed.}
\label{fig:drw_with_flares}
\end{figure*}

\begin{figure*}
\centering
\includegraphics[width=0.8\textwidth]{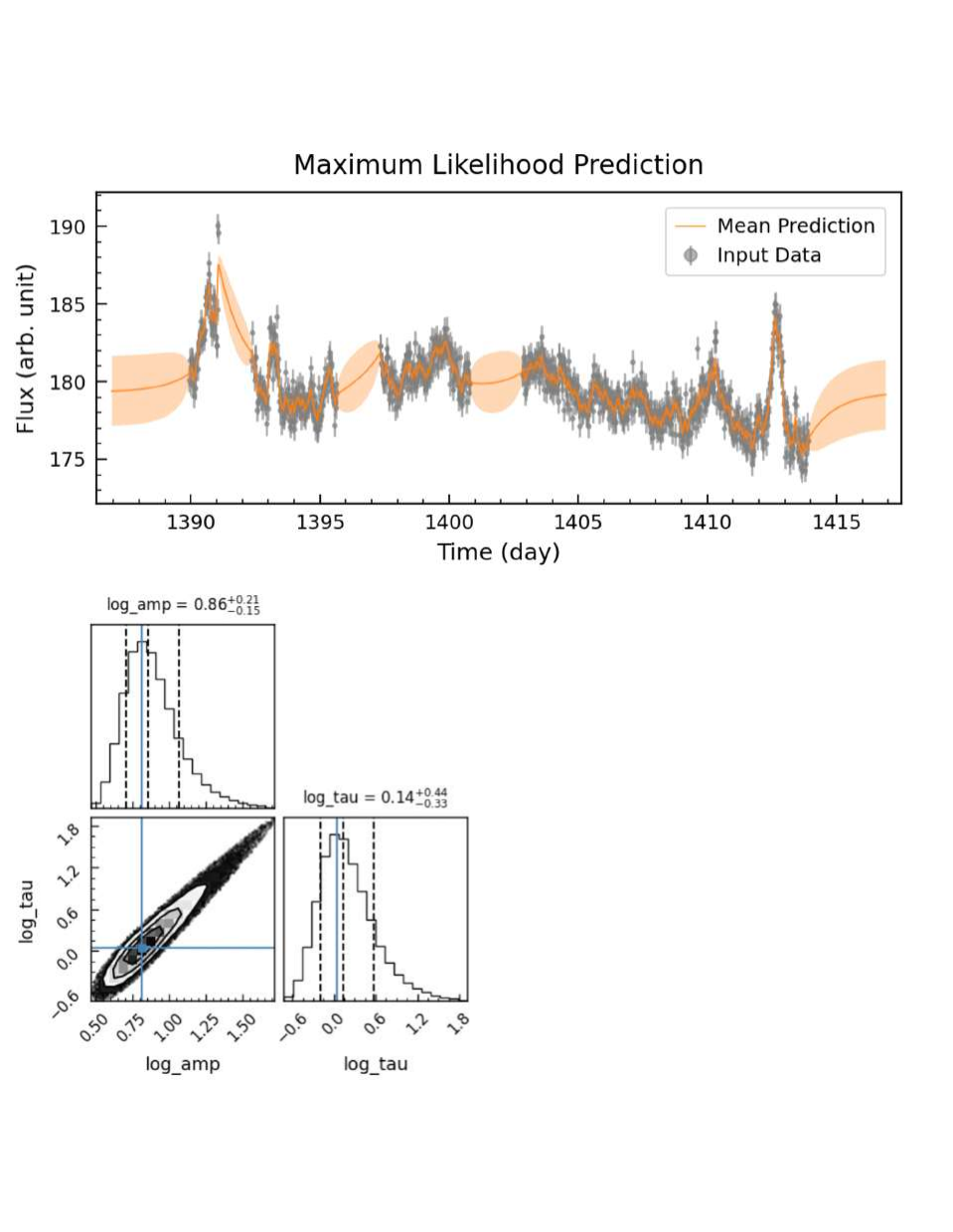}
\caption{Results of the DRW fit to Sector~5 data after the five highest-amplitude flares have been masked and removed.}
\label{fig:drw_flare_masked}
\end{figure*}

In Figures~\ref{fig:drw_with_flares} and \ref{fig:drw_flare_masked}, we show the DRW fit in the light curve of Sector~5 with and without the masking of the five highest-amplitude flares. 

\clearpage

\bibliographystyle{aasjournal}
\bibliography{BibDesk_Lia_KLS} 



\end{document}